\documentclass[aps,pre,groupedaddress]{revtex4}
\usepackage{graphicx,amsmath}
\def\*{{\bf ***}}

\def\a{\alpha}
\def\b{\beta}

\def\de{\delta}   %% NON ridefinire come \d !!!!

\def\phi{\varphi}
\def\la{\lambda}

\def\s{\sigma}

\def\om{\omega}

\def\vth{\vartheta}
\def\vphi{\varphi}

\def\R{{\bf R}}

\def\pa{\partial}

\def\d{{\rm d}}       %% derivative

\def\o+{\oplus}

\def\<{\langle}
\def\>{\rangle}

\def\interno{\hskip 2pt \vbox{\hbox{\vbox to .18
truecm{\vfill\hbox to .25 truecm
{\hfill\hfill}\vfill}\vrule}\hrule}\hskip 2 pt}

\def\({\left(}
\def\){\right)}
\def\[{\left[}
\def\]{\right]}
\def\=#1{\bar #1}
\def\~#1{\widetilde #1}
\def\.#1{\dot #1}
\def\^#1{\widehat #1}
\def\"#1{\ddot #1}

\newcommand{\eq}{\begin{equation}}
\newcommand{\feq}{\end{equation}}
\newcommand{\eqn}{\begin{eqnarray}}
\newcommand{\feqn}{\end{eqnarray}}
\newcommand{\arr}{\begin{eqnarray*}}
\newcommand{\farr}{\end{eqnarray*}}
\newcommand{\beq}{\begin{equation}}
\newcommand{\eeq}{\end{equation}}
\newcommand{\bea}{\begin{eqnarray}}
\newcommand{\eea}{\end{eqnarray}}
\newcommand{\lb}{\label}

\def\Y{Yakushevich }

\def\Dpsi{\theta^{(a)}_{i+1}-\theta^{(a)}_{i}}
\def\psip{{\psi}}
\def\psim{{\chi}}
\def\dpsip{{\Delta_n\psip}}
\def\dpsim{{\Delta_n\psim}}
\def\spsip{{{\rm S}_n\psip}}

\def\Omegap{{\xi}}
\def\Omegam{{\eta}}
\def\dOmegap{{\Delta_n\Omegap}}
\def\dOmegam{{\Delta_n\Omegam}}
\def\sOmegap{{{\rm S}_n\Omegap}}
\def\sOmegam{{{\rm S}_n\Omegam}}
\def\a{\alpha}
\def\energy#1#2#3#4{$E^{^{#1}}_{_{(#2,#3)}}=#4$}

\numberwithin{equation}{section}

\begin{document}

\title{A composite model for DNA torsion dynamics\footnote{Work
supported in part by the Italian MIUR under the program COFIN2004,
as part of the PRIN project {\it ``Mathematical Models for DNA
Dynamics ($M^2 \times D^2$)''}.}}

\author{Mariano Cadoni}
\email{mariano.cadoni@ca.infn.it} \affiliation {Dipartimento di
Fisica, Universit\`a di Cagliari and INFN, Sezione di Cagliari,
Cittadella Universitaria 09042 Monserrato, Italy}
\author{Roberto De Leo}
\email{roberto.deleo@ca.infn.it} \affiliation {Dipartimento di
Fisica, Universit\`a di Cagliari and INFN, sezione di Cagliari,
Cittadella Universitaria 09042 Monserrato, Italy}
\author{Giuseppe Gaeta}
\email{gaeta@mat.unimi.it} \affiliation {Dipartimento di
Matematica, Universit\`a di Milano, via Saldini 50, I--20133
Milano, Italy}

\begin{abstract}\noindent
DNA torsion dynamics is essential in the transcription process; a
simple model for it, in reasonable agreement with experimental
observations, has been proposed by Yakushevich (Y) and developed
by several authors; in this, the DNA subunits made of a nucleoside
and the attached nitrogen bases are described by a single degree
of freedom. In this paper we propose and investigate, both
analytically and numerically, a ``composite'' version of the Y
model, in which the nucleoside and the base are described by
separate degrees of freedom. The model proposed here contains as a
particular case the Y model and shares with it many features and
results, but represents an improvement from both the conceptual
and the phenomenological point of view. It provides a more
realistic description of DNA and possibly a justification for the
use of models which consider the DNA chain as uniform. It shows
that the existence of solitons is a generic feature of the
underlying nonlinear dynamics and is to a large extent independent
of the detailed modelling of DNA. The model we consider supports
solitonic solutions, qualitatively and quantitatively very similar
to the Y solitons, in a fully realistic range of all the physical
parameters characterizing the DNA.
\end{abstract}
\maketitle

\section{Introduction}

The possibility that nonlinear excitations -- in particular, kink
solitons or breath\-ers -- in DNA chains play a functional role
has attracted the attention of biophysicists as well as nonlinear
scientists since the pioneering paper of Englander et al.
\cite{Eng}, and the works by Davydov on solitons in biological
systems \cite{Dav}.

A number of mechanical models of the DNA double chain have been
proposed over the years, focusing on different aspects of the DNA
molecule and on different biological, physical and chemical
processes in which DNA is involved.

Here we will not discuss these, but just refer the reader to the
discussions of such attempts given in the book by Yakushevich
\cite{YakuBook} and in the review paper by Peyrard \cite{PeyNLN}
(see also the conference \cite{PeyHouches}), also for what
concerns earlier attempts which constituted the basis on which the
models considered below were first formulated.\footnote{It should
be stressed that when we speak of ``mechanical models of DNA'' we
exclude consideration of the all-important interactions between
DNA and its environment. The latter includes at least the fluid in
which DNA is immersed, and interaction with this leads to energy
exchanges; one should thus include in the equations describing DNA
dynamics both dissipation terms and random terms due to
interaction with molecules in the fluid. We will work here at a
purely mechanical level, i.e. do not consider at present these
effects. Moreover, it should be mentioned that even forgetting
dissipative and brownian motion effects, one could consider
interaction with the solvent by including effective terms in the
intrapair potential $V_p$ (see below), as done e.g. in \cite{ZC};
it has been recently shown that in the context of the
Peyrard-Bishop model this leads to a sharpening of certain
transitions \cite{Web}.} Similarly, we will not describe the
structure and functioning of DNA, but just refer e.g. to
\cite{CD,FK2,Sae}. See also \cite{PeyHouches,Pushchino} for the
role of Nonlinear Dynamics modelling in the understanding of DNA,
and \cite{Lavery,Strick} for DNA single-molecule experiments
(these were initiated about fifteen years ago \cite{SFB}, but
their range and precision has dramatically increased in recent
years; the formation of bubbles in a double-stranded DNA has been
observed in \cite{Lib}).

In recent years, two models have been extensively studied in the
Nonlinear Physics literature; these are the model by Peyrard and
Bishop \cite{PB} (and the extensions of this formulated by Dauxois
\cite{DauPLA} and later on by Barbi, Cocco, Peyrard and Ruffo
\cite{BCP,BCPR}; see also Cocco and Monasson \cite{CM}. More
recent advances are discussed in \cite{PeyNLN} and
\cite{BLPT,CuS,TPM}) and the one by Yakushevich \cite{YakPLA}; we
will refer to these as the PB and the Y models respectively.

Original versions of these models are discussed in \cite{GRPD};
they are put in perspective within a ``hierarchy'' of DNA models
in \cite{YakPRE}. An attempt to blend together the two is given in
\cite{YakSBP}; see also \cite{Joy}. Interplay between radial and
torsional degrees of freedom is considered more organically in
\cite{BCP,BCPR}.

The PB model is primarily concerned with DNA denaturation, and
describes degrees of freedom related to ``straight'' (or
``radial'') separation of the two helices which are wound together
in the DNA double helical molecule. On the other hand, the Y model
-- on which we focus in this note -- is primarily concerned with
rotational and torsional degrees of freedom of the DNA molecule,
which play a central role in the process of DNA transcription
\footnote{It is appropriate, in this context, to mention earlier
models proposed by Fedyanin, Gochev and Lisy \cite{Fedyanin}, Muto
{\it et al.} \cite{Muto}, Prohofsky \cite{Proho}, Takeno and Homma
\cite{Takeno}, van Zandt \cite{VanZ}, Yomosa \cite{Yomosa}, and
Zhang \cite{Zhang}.}.

In this model, one studies a system of nonlinear equations which
in the continuum limit reduce to a double sine-Gordon type
equation; the relevant nonlinear oscillations are kink solitons --
which are solitons in both dynamical and topological sense --
which describe the unwinding of the double helix in the
transcription region. The latter is a ``bubble'' of about 20
bases, to which RNA Polymerase (RNAP) binds in order to read the
base sequence and produce the RNA Messenger; the RNAP travels
along the DNA double chain, and so does the unwound region. The
proposal of Englander et al. \cite{Eng} was that if the nonlinear
excitations are not created or forced by the RNAP but are anyway
present due to the nonlinear dynamics of the DNA double helix
itself, a number of questions -- in particular, concerning energy
flows -- receive a simple explanation.

The Y model has been studied in a number of paper, in particular
for what concerns its solitonic solutions; here we will quote in
particular \cite{GaePLA1,GaePLA2,Gonz,YakSBP,YakPhD,YakPRE}. It
has been shown that it gives a correct prediction of quantities
related to small amplitude dynamics, such as the frequency of
small torsional oscillations; as well as of quantities related to
fully nonlinear dynamics, such as the size of solitonic
excitations describing transcription bubbles \cite{GRPD,YakuBook}.
Moreover, in its ``helicoidal'' version, it provides a scenario
for the formation of nonlinear excitation out of linear normal
modes lying at the bottom of the dispersion relation branches
\cite{GRPD}. On the other hand, if we try to fit the observed
speed of waves along the chain \cite{YakuBook}, this is possible
only upon assuming unphysical values for the coupling constants
\cite{YakPRE}.

The Y model is also a very simple one, and adopts the same kind of
simplification as in the PB model. In particular, two quite strong
features of the models are:
\begin{itemize}
\item {\it (a)} there is a single (angular) degree of freedom for
each nucleotide; \item $(b)$ all bases are considered as
identical.
\end{itemize}

These were in a sense at the basis of the success of the model, in
that thanks to these features the model can be solved exactly and
one can check that predictions allowed by the model correspond to
the real world situations for certain specific quantities. But the
features mentioned above are of course not in agreement with the
real situation.

Indeed, it is well known that bases are quite different from each
other, and in particular purines are much bigger than pirimidines;
hence feature (b) -- albeit necessary for an analytical treatment
of the model -- is definitely unrealistic.
Moreover, it is quite justified to consider several groups of
atoms within a single nucleotide (the phosphodiester chain, the
sugar ring, and the nitrogen base) as substantially rigid
subunits; but these -- in particular the sugar ring -- have some
degree of flexibility, and what's more they have a considerable
freedom of displacement -- in particular for what concerns
torsional and rotational movements -- with respect to each other.
Thus, even in a simple modelling, feature (a) is not justified
{\it per se}, and it seems quite appropriate to consider several
subunits within each nucleotide. In this sense, we will speak of a
{\it composite Yakushevich (Y) model}.
Needless to say, if such a more detailed modelling would produce
results very near to those of the simple Y model, this should be
seen as a confirmation that the latter correctly captures the
relevant features of DNA torsional dynamics -- hence justify {\it
a posteriori} feature (a) of the Y model.

In this work we propose and study a composite Y model (in the
sense mentioned above), in which we describe  with two independent
 angular degrees of freedom, the nucleoside (i.e. the segment of the
sugar-phosphate backbone pertaining to the nucleotide) and the
nitrogen base in each nucleotide.

The purpose of our study of such a model is to shed light on the
following points.

\begin{itemize}
\item {\it (A)} We want to take care (to some
extent) of correcting feature (a) above and thus investigating --
by comparing results -- how justified is the original Y modelling
in terms of one degree of freedom per nucleotide.

\item {\it (B)} We  aim at opening the way
to correct -- or justify -- feature (b) of the Y model. Indeed, in
our model we will consider separately the part of the nucleotide
which precisely replicates identical in each nucleotide (the unit
of the sugar-phosphate backbone), and the part which varies from
one nucleotide to the other (the bases). We will then be able to
study the different role of the two.

\item{\it (C)} We want to check the dependence of the solitonic
solution of the  model both from the geometry and from the value
of the physical parameters chosen. In particular, we would like to
understand how far the existence of solitons is a generic feature
of DNA and if a more realistic choice of the model geometry is
consistent with phenomenologically acceptable values of the
physical parameters.
\end{itemize}

It will turn out that the Y model, which can be considered as a
particular case of our model, captures the essential features of
DNA nonlinear dynamics. The more realistic geometry of the model
we use in this paper enables a drastic improvement of the
descriptive power of our model at both the conceptual and the
phenomenological level: on the one hand the composite Y model
keeps almost all the relevant features of the Y model, but on the
other hand it allows for more realistic choice of the physical
parameters.

It will turn out that the different degrees of freedom we use play
a fundamentally different role in the description of DNA nonlinear
dynamics. The backbone degrees of freedom are ``topological'' and
play to some extent a ``master'' role, while those associated to
the base are ``nontopological'' and are (in fluid dynamics
language) ``slaved'' to former ones. This opens an interesting
possibility, i.e. to consider a more realistic model, in which
differences among bases are properly considered, as a perturbation
of our idealized uniform model. As the essential features of the
fully nonlinear dynamics are related only to backbone degrees of
freedom, we expect that such a perturbation -- albeit with
relevant difference in the quantitative values of some parameters
entering in the model (the base dynamical and geometrical
parameters) -- will show the same kind of nonlinear dynamics as
our uniform model studied here.

The paper is organized as follows. In Sect. \ref{s2}  we will
briefly review some basic know facts about the DNA  structure and
modelling. In Sects.  \ref{s3} and \ref{s4} we will set up our
model,  describe the interaction and write down the equations of
motions that govern its dynamics. The physical parameters
characterizing our model are discussed in Sect. \ref{s5}. In Sect.
\ref{s6} we discuss the linear approximation  of our dynamical
system, in particular its dispersion relations.  In Sect. \ref{s7}
we will set up the framework for the investigation of the
nonlinear dynamics and  the topological excitations of our model.
In sect. \ref{s8}, we will show how the Y model and Y solitons
emerge as a particular case of our composite Y model and its
solitons. Sect. \ref{s9}  we investigate and derive numerically
the solitonic solutions of our model. Finally in Sect. \ref{s10}
we summarize our work and present our conclusions.

\section{DNA structure and modelling}
\lb{s2}

DNA is a gigantic polymer, made of two helices wound together; the
helices have a directionality and the two helices making a DNA
molecule run in opposite direction. We will refer for definiteness
to the standard conformation of the molecule (B-DNA); in this the
pitch of the helix corresponds to ten base pairs, and the distance
along the axis of the helix between successive base pairs is $\de
= 3.4~\AA$.

The general structure of each helix can be described as follows.
The helix is made of a sugar-phosphate backbone, to which bases
are attached. The backbone has a regular structure consisting of
repeated identical units ({\it nucleosides}); bases are attached
to a specific site on each nucleoside and are of four possible
types. These are either purines, which are adenine (A) or guanine
(G), or pyrimidines, which are cytosine (C) or thymin (T) in DNA.
It should be noted that the bases are rather rigid structures, and
have an essentially planar configuration. A unit of each helix is
called a {\it nucleotide}; this is the complex of a nucleoside and
the attached base.

The winding together of the two helices makes that to each base
site on the one helix corresponds a base site on the other helix.
Bases at corresponding sites form a base pair; each base has only
a possible partner in a base pair; bases
in a pair are linked together via hydrogen bonds (two for A-T
pairs, three for G-C pairs). The base pairs can be ``opened''
quite easily, the dissociation energy for each H-bond being of the
order of 0.04 eV, hence $\Delta E \simeq 0.1$ eV per base pair.
Opening is instrumental to a number of processes undergone by DNA,
among which notably transcription, denaturation and replication.

The backbone structure (see Fig.1)  is made of a phosphodiester
chain and a sugar ring. To one of the $C$ atoms of the sugar ring
is attached a base; this is one of the four possible bases
$A,C,G,T$, whose sequence represents the information content of
DNA and is different for different species, and to some extent for
each individual.
Thus, each helix is made of a succession of identical nucleosides,
and attached bases which can be different at each site. Bases at
corresponding sites on the two helices form a base pair, and these
can be only of two types, G-C and A-T.

\begin{figure}
\includegraphics[width=300pt]{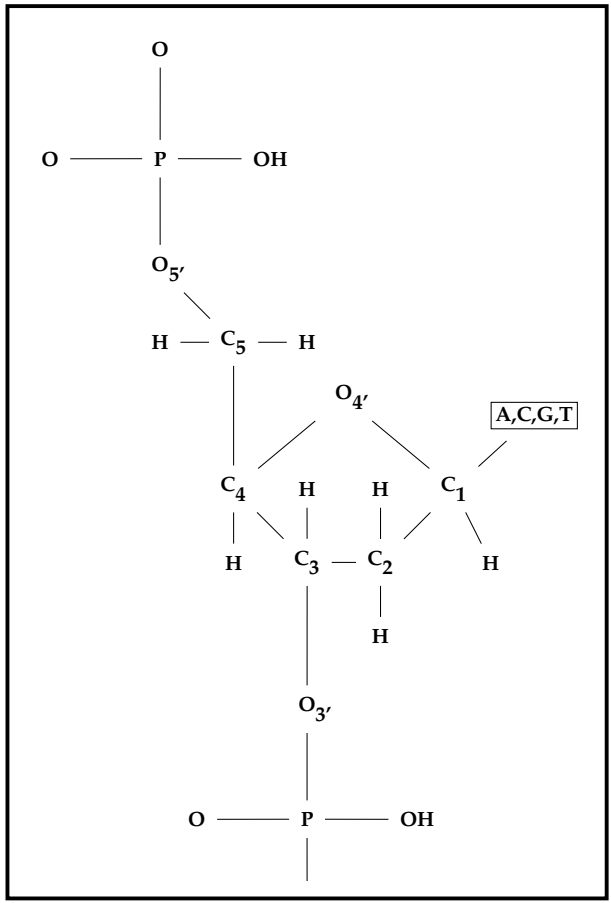}\\
  \caption{The structure of a DNA helix. A nucleotide is shown; nitrogen bases
  are attached to the $C_1$ atom in the sugar ring.}\label{figDNA}
\end{figure}

%\subsection{Interactions}

The atoms on each helix are of course held together by covalent
bonds; apart from these, other interactions should be taken into
account when attempting a description of the DNA molecule.
The backbone structure has some rigidity; in particular, it would
resist movements which represent a torsion of one nucleoside with
respect to neighboring ones. We will refer to the interaction
responsible for the forces resisting these torsion as {\it
torsional interactions}.

As already mentioned, the two bases composing a base pair are
linked together by hydrogen bonds; we will refer to the
interaction mediated by these as {\it pairing interaction}.
Each base interacts with bases at neighboring sites on the same
chain via electrostatic forces (bases are strongly polar); these
make energetically favorable the conformation in which bases are
stacked on top of each other, and therefore are referred to as
{\it stacking interaction}.

Finally, water filaments -- thus, essentially, bridges of hydrogen
bonds -- link units at different sites; these are also known as
Bernal-Fowler filaments \cite{Dav}. In particular, they have a
good probability to form between nucleosides or bases which are
half-turn of the helix apart on different chains, i.e. which are
near to each other in space due to the double helix geometry;
these water filaments-mediated interactions are therefore also
called {\it helicoidal interactions} \footnote{It is usual, for
ease of language, to refer to models in which helicoidal
interactions are taken into account as ``helicoidal'', and to
models in which they are overlooked as ``planar''. Needless to
say, the geometry of the model is the same in both cases.}.
We stress that these are quite weaker than other interactions, and
can be safely overlooked when we consider the fully nonlinear
regime. They are instead of special interest when discussing small
amplitude (low energy) dynamics, as -- just because of their
weakness -- they are easily excited and introduce a length scale
in the dispersion relations (see below).

If we consider large amplitude deviations from the equilibrium
configurations, then motions will not be completely free: the
molecule is densely packed, and the presence of the
sugar-phosphate backbone -- and of neighboring bases -- will cause
steric hindrances to the base movements. In particular, for the
rotations in a plane perpendicular to the double helix axis, the
bases will not be able to rotate around the $C_1$ atom for more
than a maximum angle $\phi_0$ without colliding with the
nucleoside.

This will lead of course to complex behaviors as the DNA helix
gets unwound; in particular, as $\phi$ gets near to its limit
value $\phi_0$ we expect some kind of essentially (if not
mathematically) discontinuous behavior. This should not be seen as
shortcoming of the model: it is indeed well known that bases
rotate in a complex way while flipping about the DNA axis (see
e.g. \cite{Banavali}).

Finally, we mention that here we consider a DNA molecule without
taking into account its macro-conformational features; that is, we
consider an ``ideal'' molecule, disregarding supercoiling,
organization in istones, and all that \cite{CD}.

\section{Composite Y model}
\lb{s3}

As mentioned above, we will model the molecule as made of
different parts (units), each of them behaving as a single
element, i.e. as a rigid body. We consider each nucleoside N as a
unit, to which a base B (considered again as a single unit) is
attached.

\subsection{General features}

We will hence model each of the helices in the DNA double chain as
an array of elements (nucleotides) made of two subunits; one of
these subunits model the nucleoside, the other the nitrogen base.
We will consider the bases as all equal, thus disregarding the
substantial difference between them \footnote{This assumption (see
also section 9) is common to all the mathematical -- as opposed to
physico-chemical -- models of DNA, and as already remarked is
necessary to be able to perform an analytical study of the model.
We refer e.g. to \cite{PeyNLN,YakuBook} for discussion about this
point. Study of real sequences, i.e. with different
characteristics for different bases, is possible numerically; see
e.g. \cite{Sal,ZC}.}. The chains -- and thus the arrays -- will be
considered as infinite.

We will use a superscript $a=1,2$ to distinguish elements on the
two chains, and a subscript $i \in {\bf Z}$ to identify the site
on the chains. Thus the base pairing will be between bases
$B_i^{(1)}$ and $B_i^{(2)}$, while stacking interaction will be
between base $B_i^{(a)}$ and bases $B_{i+1}^{(a)}$ and
$B_{i-1}^{(a)}$.

We will consider each nucleoside $N^{(a)}_i$ as a disk; bases will
be seen as disks themselves, with a point on the border of
$B^{(a)}_i$ attached via an inextensible rod to a point $p_c$ on
the border of $N^{(a)}_i$; these points on B and N represent the
locations of the $N$ atom on B and of the $C_1$ atom on N involved
in the chemical bond attaching the base to the nucleoside. The rod
can rotate by an angle $\pm \phi_0$ before B collides with N; on
the other hand, the disk N can rotate completely around its axis.

\begin{figure}
\includegraphics[width=300pt,bb = 110 610 502 762]{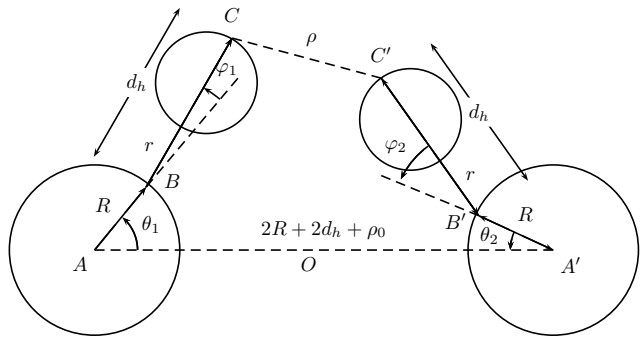}\\
  \caption{A base pair in our model. The origin of the coordinate system is in O.
  The angles $\theta_1$ between the lines AO and AB and $\theta_2$
  between A'O and A'B' correspond
  to torsion of sugar-phosphate backbone with respect to the equilibrium
  B-DNA conformation; the angles $\phi_1$ between the line AB and the line BC, and
  $\phi_2$ between A'B' and B'C' correspond to rotation of bases around
  the $C_1 - N$ bond linking them to the nucleotide. All angles are in counterclockwise
  direction; thus the angles $\theta_2$ and $\phi_2$ in the figure are negative.}
  \label{modello1}
\end{figure}

We also single out a point $p_h$ on the border of the disk
modelling the base; this represents the atom(s) which form the H
bond with the corresponding base on the other DNA chain.

The disks (i.e. the elements of our model) are subject to
different kinds of forces, corresponding to those described above:
torsional forces resisting the rotation of one disk $N^{(a)}_i$
with respect to neighboring disks $N^{(a)}_{i+1}$ and
$N^{(a)}_{i-1}$ on the same chain; stacking forces between a base
$B_i^{(a)}$ and neighboring bases $B_{i+1}^{(a)}$ and
$B_{i-1}^{(a)}$ on the same chain; pairing forces between bases
$B_i^{(1)}$ and $B_i^{(2)}$ in the same base pair; and finally,
helicoidal forces correspond to hydrogen bonded Bernal-Fowler
filaments linking bases $B_i^{(1)}$ and $B_{i\pm 5}^{(2)}$ (and
$B_i^{(2)}$ and $B_{i\pm 5}^{(1)}$).

\subsection{The Lagrangian}

We should now translate the above discussion into a Lagrangian
defining our model. This will be written as
\beq\lb{lag} L \ = \ T
\ - \ \( U_t \ + \ U_s \ + \ U_p \ + \ U_h \) \feq
where $T$ is
the kinetic energy, and $U_a$ are the potential energies for the
different interactions listed above, i.e.
\begin{itemize}
\item $U_t$ is the backbone torsional potential, \item $U_s$ is
the stacking potential, \item $U_p$ is the pairing potential,
\item $U_h$ is the helicoidal potential. \end{itemize}
These will be modelled by two-body potentials, for which we use
the notation $V_a$, to be summed over all interacting pairs in
order to produce the $U_a$.

We denote by $I$ the moment of inertia (around center of mass) of
disks modelling the nucleosides, and by $I_B$ the moment of
inertia of bases around the $C_1$ atom in the sugar ring; as the
bases can not rotate around their center of mass, this is equal to
$m r^2$, where $m$ is the base mass and $r$ is the distance
between the $C_1$ atom in the sugar and the center of mass of the
base.

\subsection{The degrees of freedom}

We are primarily interested in the torsional dynamics. Thus, for
each element we will consider torsional movements, hence a
rotation angle (with respect to the equilibrium conformation,
which we take for definiteness to be B-DNA); these will be denoted
as $\theta^{(a)}_i$ for the nucleoside $N^{(a)}_i$, and
$\phi^{(a)}_i$ for the base $B^{(a)}_i$. Only these rotations will
be allowed in our model. All angles will be positive in
counterclockwise sense.

The angles $\theta$ represent a torsion of the sugar-phosphate
backbone with respect to the equilibrium configuration; thus they
are related to unwinding of the double helix. On the other hand,
the angles $\phi$ represent a rotation of the base with respect to
the corresponding nucleoside; the motion described by $\phi$ can
be thought of as a rotation around the $C_1$ atom in the sugar
ring. Note that the hindrances due to the presence of backbone
atoms constrain rotation of the base around the $C_1$ atom.

Thus, as mentioned above, the angles will have a different range
of values: \beq\lb{bc} \theta^{(a)}_i \in \R \ , \ \ \phi^{(a)}_i
\in [- \phi_0,\phi_0] \ \ \phi_0 < \pi \ . \feq
The actual value of $\phi_0$ is not essential. The important
feature is that the base can not rotate freely around the $C_1$
atom, but only pivot between certain limits. At the level of the
numerical analysis the simplest way to implement the previous
boundary condition is to use a confining potential, which
reproduces  approximately the form of a box. For this reason in
Sect. \ref{s9} we will add to the Hamiltonian of the system a
confining potential $V_{w}= K \tan^{4}\phi^{(a)}$.

It should be stressed that -- just on the basis of these different
ranges of variations -- there will be a substantial difference
between the degrees of freedom described by the angles: those
described by $\theta$ angles will be {\it topological} degrees of
freedom, while those described by $\phi$ angles will only describe
local and (relatively) small motions -- hence $\phi$ describe {\it
non topological} degrees of freedom.

\subsection{Cartesian coordinates}

In computing the kinetic energy, it will be convenient to consider
cartesian coordinates. With reference to Fig. \ref{modello1}, the cartesian
coordinates in the $(x,y)$ plane orthogonal to the double helix
axis of relevant points will be as follow.

The center of disks, representing the position of the
phosphodiester chain, will be $(x^{(a)}_o , y^{(a)}_o)$; the point
on the border of the disks representing the $C_1$ atom to which
the base disks are attached will be $(x^{(a)}_c , y^{(a)}_c)$. The
center of mass of the bases will be $(x^{(a)}_b , y^{(a)}_b)$, and
the point on the border of the disks modelling bases representing
the atom(s) forming the H bonds will be $(x^{(a)}_h , y^{(a)}_h)$.

In terms of the $\{ \theta , \phi \}$ angles, these are given by
(we omit the site index $i$ for ease of writing, and give
condensed formulas for the two chains, with first sign referring
to chain 1): \beq\lb{cartesian} \begin{array}{ll}
x^{(1,2)}_o = \mp a , & y^{(1,2)}_o = 0 ; \\
x^{(1,2)}_c = x^{(1,2)}_o \pm R \cos ( \theta^{(1,2)} ) , &
y^{(1,2)}_c = \pm R \sin ( \theta^{(1,2)} ) ; \\
x^{(1,2)}_b = x^{(1,2)}_c \pm r \cos ( \theta^{(1,2)} +
\phi^{(1,2)} ) , & y^{(1,2)}_b = y^{(1,2)}_c \pm r \sin (
\theta^{(1,2)} + \phi^{(1,2)} ) ; \\
x^{(1,2)}_h = x^{(1,2)}_c \pm d_h \cos ( \theta^{(1,2)} +
\phi^{(1,2)} ) , & y^{(1,2)}_h = y^{(1,2)}_c \pm d_h \sin (
\theta^{(1,2)} + \phi^{(1,2)} ). \\
\end{array} \feq
Here and in the following we denote by $R$ the radius of disks
describing nucleosides, i.e. the length of the segments AB and
A'B' in Fig. \ref{modello1} (this is the distance from the
phosphodiester chain to the $C_1$ atom); by $r$ the distance
between the center of mass of bases and the border of the disk
modelling the nucleoside (i.e. the $C_1$ atom). We also denote by
$d_h$ the lengths (supposed equal) of the segments BC and B'C'
joining the $C_1$ atom on the nucleoside and the atoms of the
bases forming the hydrogen bond linking this to the complementary
base. The parameter $a$ corresponds to the distance between the
double helix axis and the phosphodiester chain, whereas
$\rho_{0}$ is the distance between points $C$ and $C'$ in the
equilibrium configuration. The previous parameters are obviously
related by the equation $2a=2R+2d_{h}+\rho_{0}$.

\subsection{Kinetic energy}

With this notation, and standard computations, the kinetic energy
of each nucleotide is written as
$$ T_i^{(a)} = \frac {1}{2}
\[ m  r^2 {\dot\phi}^2 + 2  m  r  ( r + R \cos \phi )
{\dot\theta}  {\dot\phi} + \( I + m_B (R^2 + r^2 ) + 2 m R r
 \cos \phi \) {\dot\theta}^2 \] , $$
 where we have suppressed super- and sub-scripts for ease of
 reading. Thus, the total kinetic energy for the double chain is
\beq\lb{kin} \begin{array}{rl}
T \ =& \ \sum_a \ \sum_i \ T^{(a)}_i \ = \\
 =& \ \frac {1} {2} \, \sum_a \sum_i
 \ \[ m \, r^2 \, \( {\dot\phi}^{(a)}_i \)^2 \ + \ 2 \, m \, r \,
 ( r + R \cos (\phi^{(a)}_i) ) \,
{\dot\theta}^{(a)}_i \, {\dot\phi}^{(a)}_i \ + \right. \\
 & \left. + \ \( I + m_B (R^2 + r^2 ) + 2 m R r
 \cos (\phi^{(a)}_i) \) \( {\dot\theta}^{(a)}_i \)^2 \] \  \ .
\end{array} \feq

We have thus considered a general class of composite Y models; so
far we have not specified the interaction potentials, which are
needed to have a definite model.

\subsection{Modelling the interactions}

We have now to specify our model by fixing analytical expressions
for terms modelling potential interactions in the lagrangian
(\ref{lag}).
As one of our aims is to compare the results obtained by a
composite Y model with those obtained with the simple Y model, we
will make choices with the same physical content as those made by
Yakushevich.

\subsubsection{Torsional interactions}

Torsional forces will depend only on difference of angles
(measured with respect to the equilibrium B-DNA configuration) of
neighboring units on the same phosphodiester chain; thus we
introduce a torsion potential $V_t$ and have
\beq\lb{tp}
U_t \ = \ \sum_a
\sum_i V_t \( \theta^{(a)}_{i+1} - \theta^{(a)}_i \) \ . \feq
The potential $V_t$ must have a minimum in zero, and be $2
\pi$-periodic in order to take into account the fundamentally
discrete and quantum nature of the phosphodiester chain. Here we
will take the simplest such function \footnote{A more realistic
choice could have important consequences of qualitative -- and not
just quantitative -- behavior of nonlinear excitations. This point
is discussed in \cite{SacSgu}.}, i.e. (adding an inessential
constant so that the minimum corresponds to zero energy) \beq V_t
(x) \ = \ K_t \( 1 \ - \ \cos (x) \) \feq where $K_t$ is a
dimensional constant.
Thus, our choice for torsional interactions will be
\beq\lb{Ut}
U_t \ = \ K_t \ \sum_a \sum_i \, \[ 1 \, - \, \cos \(
\theta^{(a)}_{i+1} - \theta^{(a)}_i \) \] \ . \feq
The harmonic
approximation for this is of course $$ U_t^q \ = \frac{1}{2} K_t \
\sum_a \sum_i \, \( \theta^{(a)}_{i+1} - \theta^{(a)}_i \)^2  \ .
$$

\subsubsection{Stacking interactions}

Stacking between bases will only depend on the relative
displacement of neighboring bases on the same helix in the plane
orthogonal to the double helix axis \footnote{The stacking
interactions are of essentially electrostatic nature; thus it is
reasonable in this context to see the bases as dipoles. If we have
two identical dipoles made of charges $\pm \alpha$ a distance $d$
apart, their separation vector being along the $z$ direction, and
force them to move in parallel planes orthogonal to the $z$ axis
and a distance $L$ apart, then denoting with $\s$ the distance of
their projections in the $(x,y)$ plane we have
$$ V (\rho ) = (1/2) \a ((d+L)^{-3} - 2 L^{-3} + (d-L)^{-3})\s^2 - (3/8)
\a ((d+L)^{-5} - 2 L^{-5} + (d-L)^{-5}) \s^4 + O (\s^6) \ .
$$}. That is, introducing a stacking potential $V_s$ we have
\beq\lb{Us1} U_s \ = \ \sum_a
\sum_i V_s \( \s^{(a)}_{i} \) \ , \feq where \beq \s^{(a)}_i \ :=
\ \sqrt{(x^{(a)}_{i+1} - x^{(a)}_i)^2 \ + \ (y^{(a)}_{i+1} -
y^{(a)}_i)^2} \ , \feq
where $x^{(a)}_i,\,y^{(a)}_i$ are the coordinates of the center of
mass of the bases.
The simplest choice corresponds to a harmonic potential
\footnote{The interaction does more properly depend on the degree
of superposition of projections to bases on the plane orthogonal
to the double helix axis (and moreover depends on the details of
charge distribution on each base), and in particular quickly goes
to zero once the bases assume different positions. Moreover, once
the bases extrude from the double helix there are ionic
interactions between the bases and the solvent which should be
taken into account \cite{ZC}. In this note, however, we will just
consider harmonic stacking.}, $V_s = (1/2) K_s \s^2$. This will be
our choice -- which again corresponds to the one made in the PB
and in the Y models, so that
\beq U_s \ = \ \sum_a \sum_i
\frac{K_s}{ 2} \, (\s^{(a)}_{i} )^2 \ . \feq
We should however express this in terms of the $\theta$ and $\phi$
angles. With standard algebra, using Eqs. (\ref{cartesian}),  we obtain \beq\lb{Us}
\begin{array}{rl} U_s \ =& \ \frac{1}{2} \, K_s \ \sum_a \sum_i \
2 \, \[ R^2 +
r^2 \ + \right. \\
& \left. - \ R^2 \cos (\theta^{(a)}_{i+1} - \theta^{(a)}_{i} ) \,
- \, r^2 \cos [ (\theta^{(a)}_{i+1} - \theta^{(a)}_{i} ) +
(\phi^{(a)}_{i+1} - \phi^{(a)}_{i}) ] \ + \right. \\
 & \left. - \ R r \( \cos [(\theta^{(a)}_{i+1} - \theta^{(a)}_{i})
 + \phi^{(a)}_{i+1} ] + \cos [(\theta^{(a)}_{i+1} - \theta^{(a)}_{i})
 - \phi^{(a)}_{i} ] \) \ + \right. \\
  & \left. + \ R r \( \cos (\phi^{(a)}_{i+1}) + \cos(\phi^{(a)}_{i})
 \) \] \ . \end{array} \feq

\subsubsection{Pairing interactions}

Pairing interactions are due to stretching of the hydrogen bonds
linking bases in a pair. Introducing a paring potential $V_p$
which models the H bonds, we have
\beq U_p \ = \ \sum_i \, V_p
(\theta^{(1)}_{i} , \theta^{(2)}_{i}, \phi^{(1)}_{i},
\phi^{(2)}_{i}) \ . \feq
We note that H bonds are strongly directional, so that they are
quickly disrupted once the alignment between pairing bases is
disrupted. This feature is traditionally disregarded in the Y
model, where it is assumed that $V_p$ only depends on the distance
\beq \rho_i \ := \sqrt{ ( x^{(1)}_{i} - x^{(2)}_{i} )^2  + (
y^{(1)}_{i} - y^{(2)}_{i} )^2} \feq between the interacting bases;
that is, \beq\lb{Up1} U_p \ = \ \sum_i \, V_p (\rho_{i} ) \ . \feq
As noted by Gonzalez and Martin-Landrove
\cite{Gonz} in the context of the Yakushevich model, one should be
careful in expanding a potential $V_p (\rho)$ in terms of the
rotation angles $\phi$ and $\theta$: indeed, unless $\rho_{0}=0$, i.e.
$a = R + d_h$,
one would get zero quadratic term in such an expansion (see
however \cite{GaeY1} for what concerns solitons in this context).

As for the potential $V_p$, there are two simple choices for this
appearing in the literature. On the one hand, Yakushevich
\cite{YakPLA} suggests to consider a potential harmonic in the
intrapair distance $\rho$ (this would appear nonlinear when
expressed through rotation angles) and this has been kept in
subsequent discussions and extensions of her model
\cite{YakuBook}; on the other hand, Peyrard and Bishop \cite{PB}
consider a Morse potential; again this has been kept in subsequent
discussions and extensions of their model \cite{PeyNLN}.

There is no doubt that the Morse potential is more justified in
physical terms; however, as we wish to compare our results with
those of the original Y model, we will at first
consider a harmonic
potential
\beq\lb{Vp} V_p^{(Y)} (\rho) \ = \ \frac{1}{2} \, K_p \,
(\rho - \rho_0 )^2 \ , \feq
where $\rho_0$ is the intrapair distance in the
equilibrium configuration. Moreover, again in order to compare our
results with those of the original Y model, we will later on set
$\rho_0 = 0$. This corresponds to setting $a = R + d_h$. These
approximations can appear very crude, but experience gained (as
preliminary work for the present investigation) with the standard
\Y model \cite{GaeY1,GaeY2} suggests they do not have a great
impact at the level of fully nonlinear dynamics.

We should express $V_p$ in terms of the rotations angles. Using
once again the expressions (\ref{cartesian}), we have with
standard computations that
\beq\lb{rho} \begin{array}{rl} \rho_i^2
\ :=& \( x^{(1)}_{i} - x^{(2)}_{i} \)^2 \ + \ \(
y^{(1)}_{i} - y^{(2)}_{i} \)^2 \ = \\
 =&
2 \[ 2 a^2 + R^2 + d_{h}^2  +
    R^2 \cos (\theta^{(1)}_i  - \theta^{(2)}_i )  +
    d_{h}^2 \cos [(\theta^{(1)}_i  - \theta^{(2)}_i ) +
    (\phi^{(1)}_i  - \phi^{(2)}_i ) + \right. \\
 & \left. + \ R d_{h} \( \cos \phi^{(1)}_i  + \cos \phi^{(2)}_i +
    \cos [(\theta^{(1)}_i  - \theta^{(2)}_i ) + \phi^{(1)}_i  ] +
    \cos [(\theta^{(1)}_i  - \theta^{(2)}_i ) - \phi^{(2)}_i  ]  \) + \right. \\
 & \left. - 2 a R \( \cos (\theta^{(1)}_i ) + \cos (\theta^{(2)}_i
 )\) - 2 a d_{h}\(
    \cos (\phi^{(1)}_i  + \theta^{(1)}_i )  +
    \cos (\phi^{(2)}_i  + \theta^{(2)}_i ) \) \] \ . \end{array} \feq
With this, our choice for the pairing part of the hamiltonian will
be \beq\lb{Up} U_p \ = \ \sum_i \, V_p (\rho_i ) \ . \feq

\subsubsection{Helicoidal interactions}

Helicoidal interaction are mediated by water filaments
(Bernal-Fowler filaments \cite{Dav}) connecting different
nucleotides; in particular we will consider those being on
opposite helices at half-pitch distance, as they are near enough
in three-dimensional space due to the double helical geometry. As
the nucleotide move, the hydrogen bonds in these filaments -- and
those connecting the filaments to the nucleotides -- are stretched
and thus resist differential motions of the two connected
nucleotides.

We will, for the sake of simplicity and also in view of the small
energies involved, only consider filaments forming between
nucleosides; thus only the $\theta$ angles will be involved in
these interactions.
We have therefore, introducing a helicoidal potential $V_h$ and
recalling that the pitch of the helix corresponds to 10 bases in
the B-DNA equilibrium configuration, \beq\lb{Uh1} U_h \ = \ \sum_i
\, V_h (\theta^{(1)}_{i+5} - \theta^{(2)}_{i}) \ + \ V_h
(\theta^{(2)}_{i+5} - \theta^{(1)}_{i}) \ . \feq As the angles
$\theta$ are involved, the potential $V_h$ should be $2
\pi$-periodic \footnote{In physical terms, this is not obvious by
itself, as the filaments could have to wind around the double
helix if the two connected nucleosides are twisted by $2 \pi$ with
respect to each other; however, when this happens the filaments
are actually broken and then built again thanks to quantum
fluctuations.}.

Such water filament connections involve a large number (around 10)
of hydrogen bonds; hence each of them is only slightly stretched,
and it makes sense to consider the angular-harmonic approximation
\beq V_h (\tau ) \ = \ K_h [1 - \cos (\tau) ] \ \simeq \ \frac{1}{2} K_h
\tau^2  \ . \feq

Our choice will therefore be \beq\lb{Uh} U_h \ = \ K_h \, \sum_i
\, \[ 2 - \cos(\theta^{(1)}_{i+5} - \theta^{(2)}_{i}) -
\cos(\theta^{(2)}_{i+5} - \theta^{(1)}_{i}) \] \ . \feq

\section{Equations of motion}
\lb{s4}

In  the previous sections we have set up the  model and the
interactions. Let us now study its dynamics. We denote
collectively the variables as $\psi^a$, e.g. with $\psi =
(\phi^{(1)},\phi^{(2)},\theta^{(1)},\theta^{(2)} )$. The dynamics
of the model will be described by the Euler-Lagrange equations
corresponding to the Lagrangian (\ref{lag}) with the terms in the
interaction potential given respectively by Eqs. (\ref{Ut}),
(\ref{Us}), (\ref{Up}), (\ref{Uh}) \beq \frac{\pa L}{\pa \psi^a_i}
\ - \ \frac{\d } {\d t} \frac{\pa L} { \pa {\dot\psi}^a_i} \ = \ 0
\ . \feq With our choices for the different terms of $L$, and
writing $\hat a$ for the complementary chain of the chain $a$ (that
is, $\hat 1 = 2$, $\hat 2=1$), these read
\beq\lb{EuLag}
\begin{array}{l}
m r^2 {\ddot \phi}^{(a)}_i + m r [R \cos (\phi^{(a)}_i) + r]
{\ddot \theta}^{(a)}_i  + m r R \sin ( \phi^{(a)}_i ) ({\dot
\theta}^{(a)}_i )^2 \ = \\
\ = \ K_s r^2 \sin[\phi^{(a)}_{i-1} - \phi^{(a)}_{i} +
\theta^{(a)}_{i-1} - \theta^{(a)}_{i}] -
  2 a d_h K_p \sin (\phi^{(a)}_{i} + \theta^{(a)}_{i} )\\ -
  K_s r R \sin (\phi^{(a)}_{i} - \theta^{(a)}_{i-1} + \theta^{(a)}_{i} ) -
  K_s r R \sin ( \phi^{(a)}_{i} + \theta^{(a)}_{i} - \theta^{(a)}_{i+1} ) \\
  \ \ \ -
  K_s r^2 \sin ( \phi^{(a)}_{i} - \phi^{(a)}_{i+1} + \theta^{(a)}_{i}
  - \theta^{(a)}_{i+1} ) +
  d_h K_p R \sin ( \phi^{(a)}_{i} + \theta^{(a)}_{i} -
  \theta^{(\hat a)}_{i} ) +\\
  d_h^2 K_p \sin ( \phi^{(a)}_{i} - \phi^{(\hat a)}_{i} +
  \theta^{(a)}_{i} - \theta^{(\hat a)}_{i} ) +
  R (d_h K_p + 2 K_s r ) \sin[ \phi^{(a)}_{i} ] \ ; \\
  {} \\
m r R \cos ( \phi^{(a)}_{i} ) ({\ddot \phi}^{(a)}_{i} + 2 {\ddot
\theta}^{(a)}_{i} ) + m r^2 {\ddot \phi}^{(a)}_{i} + I {\ddot
\theta}^{(a)}_{i} +
  m r^2 {\ddot \theta}^{(a)}_{i} + m R^2 {\ddot \theta}^{(a)}_{i}\\
  -m r R \sin (\phi^{(a)}_{i}) {\dot \phi}^{(a)}_{i} ({\dot \phi}^{(a)}_{i} +
  2 {\dot \theta}^{(a)}_{i}) \ = \\
\ = \ (K_t + K_s R^2 ) \sin ( \theta^{(a)}_{i-1} -
\theta^{(a)}_{i} )
  + K_s r R \sin ( \phi^{(a)}_{i-1} - (\theta^{(a)}_{i} -
  \theta^{(a)}_{i-1}))\\
  -  K_s r^{2} \sin ( (\phi^{(a)}_{i} - \phi^{(a)}_{i-1})
  \ \ \ + (\theta^{(a)}_{i} - \theta^{(a)}_{i-1}) ) -
  2 a K_p R \sin (\theta^{(a)}_{i}) -\\
  2 a d_h K_p \sin ( \phi^{(a)}_{i} + \theta^{(a)}_{i} )
  -K_s r R \sin [ \phi^{(a)}_{i} + (\theta^{(a)}_{i} - \theta^{(a)}_{i-1} )
  ]  \\- K_{s} r R \sin (\phi^{(a)}_{i} - (\theta^{(a)}_{i+1} -
  \theta^{(a)}_{i}) )
  + (K_t + K_s R^2) \sin (\theta^{(a)}_{i+1} - \theta^{(a)}_{i} )\\
  + K_s r^2 \sin [(\phi^{(a)}_{i+1} - \phi^{(a)}_{i}) +
        (\theta^{(a)}_{i+1} - \theta^{(a)}_{i}) ]
   + K_s r R \sin [ \phi^{(a)}_{i+1} + (\theta^{(a)}_{i+1} -
  \theta^{(a)}_{i}) ] +\\
  K_p R^2 \sin ( \theta^{(a)}_{i} - \theta^{(\hat a)}_{i})  +
  d_h K_p R \sin (\phi^{(a)}_{i} + (\theta^{(a)}_{i} - \theta^{(\hat a)}_{i}) )
  \ \ \ +\\
  d_h^2 K_p \sin ((\phi^{(a)}_{i} - \phi^{(\hat a)}_{i}) +
  (\theta^{(a)}_{i} - \theta^{(\hat a)}_i ) ) -
  d_h K_p R \sin ( \phi^{(\hat a)}_{i} - (\theta^{(a)}_{i} -
  \theta^{(\hat a)}_{i}) )\\
    + K_h \( \theta^{(\hat a)}_{i+5} -
  2  \theta^{(a)}_{i} + \theta^{(\hat a)}_{i-5} \)
\end{array} \feq
Note that here $a,R,r,d_h$ are considered as independent
parameters, i.e. we have not enforced the Yakushevich condition $R
+ d_h = a$ (i.e. $\rho_0 = 0$).
Needless to say, these are far too complex to be analyzed
directly, and we will need to introduce various kinds of
approximation.

We have thus completely specified the model we are going to study and
 derived the equations that govern its dynamics,
i.e. its lagrangian and the equations of motion. The choice of
torsion angles as variables to describe our dynamics led to
involved expressions, but our choices are very simple physically.

We have considered ``angular harmonic'' approximations (expansion up to
first Fourier mode) i.e. potentials of the form $V (x) = [1 - \cos
(x)]$ for the torsion and helicoidal interactions, harmonic
approximation for the base stacking interaction, and a harmonic
potential depending on the intrapair distance for the pairing
interaction.
Our approximations are coherent with those considered in the
literature when dealing with uniform models of the DNA chain, and
in particular when dealing with (extensions of) the Yakushevich
model.
Thus, when comparing the characteristic of our model with those of
these other models, we are really focusing on the differences
arising from considering separately the nucleoside and the base
within each nucleotide.

It would of course be possible to consider more realistic
expressions for the potentials; but we believe that at the present
stage this would rather obscure the relevant point here, i.e. the
discussion of how such ``composite'' models can retain the
remarkable good features of the Y model and at the same time
overcome some of the difficulties they encounter.
Finally, we note that it is quite obvious that the dynamical
equations describing the model are -- despite the simplifying
assumptions we made at various stage -- too hard to have any hope
to obtain a general solution, either in the discrete or in the
continuum version (see below) of the model.

In the next section we will   focus our attention on
the choice of the physical parameters appearing in our model.
Later,  we
will investigate the dynamics beginning with  the linear
approximation and then in the fully nonlinear regime.

\section{Physical values of parameters}
\lb{s5}

In order to have a well defined model
we should still assign concrete values to the
parameters -- both geometrical ones and coupling constants --
appearing in our Lagrangian (\ref{lag})  and in the  equation of
motion (\ref{EuLag}).

\subsection{Kinematical parameters}

Let us start by discussing kinematical parameters; in these we
include the geometrical parameters as well as the mass $m$ and the
moment of inertia $I$.

The masses can be readily evaluated by considering the chemical
structure of the bases. They can be calculated just by knowing
masses of the atoms and their multiplicity in the different bases.
As for the geometrical parameters like $R$, $a$, $r$ and $d_h$ (and the
moment of inertia $I$), quite surprisingly different authors seem
to provide different values for these. Rather than assuming the values
given by one or another author, we have preferred to estimate the
parameters using the available information about the DNA structure.
Position of atoms within the bases (which of course determine
$R$, $a$, $r$ and $d_h$, and hence $I$) and  geometrical descriptions of
DNA are widely available to the scientific community in form of
{\tt PDB} files \cite{PDBRep}. We will use this information
(which we accessed at \cite{1BNA} and \cite{PDBFiles}) to estimate
directly all static parameters in play on the basis of the atomic
positions.

The geometrical parameters which are relevant for our discussion
are the longitudinal width of bases  $l_{b}$ and of the sugar
$l_{s}$, the  distances of the bases from the relative sugars
$d_s$ and the distance of a base from  the relative dual base
$d_b$. We give our estimates for the masses, moments of inertia
and the parameters $l,d_{s}, d_{b}$ for the different bases and
their mean values in Table~\ref{tab:cynParms}.
>From those data and using the equations $ R = l_{s},\,
r = {\hat d}_{s}+ {\hat l}_{b}/2,\,
d_h =  {\hat l}_{b} + {\hat d}_{s},\,
a = l_{s} +
{\hat l}_{b} + {\hat d}_{s} + {\hat d}_{b} / 2$ (hats denote  mean
values),
one obtains  the average values for
the geometrical parameters appearing in our Lagrangian.
given in table \ref{tab:para}.

\begin{table}[ht]
\centering
 \begin{tabular}{|l|c|c|c|c|c|c|}
  \hline
    & A & T & G & C & mean & Sugar\\
    \hline
  $m$ & 134  & 125  & 150  & 110  & 130 & 85 \\
  $I$ & $3.6\times10^3$ & $3.0\times10^3$ & $4.4\times10^3$ & $2.3\times10^3$ & $3.3\times10^3$ & $1.2\times10^2$\\
  $l$ & $3.2$ & $4.0$ & $5.0$ & $2.4$ & $4.7$ & $3.3$\\
  $d_s$ & $1.5$ & $1.5$ & $1.5$ & $1.5$ & $1.5$ & -\\
  $d_b$ & $2.0$ & $2.0$ & $2.0$ & $2.0$ & $2.0$ & -\\
 \hline
\end{tabular}
\caption{
  Order of magnitude for the basic geometrical parameters of the DNA.
  Units of measure are: atomic unit for masses $m$, $1.67 \times 10^{-47}
  {\rm Kg}\cdot {\rm m}^2$
  for the inertia momenta $I$, Angstrom for $l$, $d_s$ and $d_b$, respectively the
  longitudinal width of bases and their distances from the relative sugars and
  from the relative dual base. These values have
  been extracted from the sample ``generic'' B-DNA PDB data~\cite{PDBFiles}, kindly
  provided by the Glactone Project~\cite{Glac}, and double checked with the data from
  \cite{1BNA}, that agree within $5\%$.
  Inertia momenta of bases has been evaluated with respect to
  rotations about the DNA's symmetry axis passing through the sugar's $C_1$ atom
  the base is attached to; the inertia momentum of the sugar itself has been evaluated
  with respect to rotations about its $C_3-C_4$ axis (see Fig.~\ref{figDNA})}
 \label{tab:cynParms}
\end{table}

\begin{table}[ht]
\centering
   \begin{tabular}{|c|c|c|c|}
    \hline
    $R$ & $r$ & $d_{h}¥$ & $a$ \\
    \hline
    3.3~\AA  &3.8~\AA  & 6.2~\AA & 10.5~\AA  \\
    \hline
   \end{tabular}
   \caption{Numerical values of the geometrical parameters
   chracterizing our model}
   \label{tab:para}
\end{table}

\subsection{Coupling constants}

The determination of the four coupling constants appearing in our
model is more
problematic, due partly to the difficulties in making experiments
to test single coupling constants and partly to the complexity of
the system itself.

\subsubsection{Pairing}

The  coupling constant $K_p$, which appears in the pairing potential
(\ref{Up}) can be easily determine by considering the typical energy
of hydrogen bonds. The pairing interaction involves
two (in the $A-T$ case) or three (in the $G-C$ case) electrostatic
hydrogen bond. The pairing potential can be modelled
with a Morse function
\beq\lb{morse}
V_p(x)=D(e^{-b d(x,x_0)}-1)^2=\frac{1}{2}(2D b^2)(\rho-\rho_0)^2+O(\rho^3)\,,
\feq
where $D$ is the potential depth, $\rho$ the distance from the
equilibrium position $\rho_0$ and $b$ a parameter that defines
the width of the well. Although throughout this paper we  use
the harmonic potential (\ref{Up}) to model the pairing interaction,
the use of the Morse function seems more appropriate for evaluating
the parameter  $K_p$. The point is that the pairing coupling constant
is physically determined by the behavior of the pairing potential
away from its minimum. Using the harmonic approximation (\ref{Up})
for estimate $K_p$ would result in a completely unphysical value for
the parameter.

Different estimates  of the parameters appearing in the potential
(\ref{morse}) are present in the literature.
The estimates
$$ D_{AT} = 0.030 {\rm eV} , \ D_{GC} = 0.045 {\rm eV}, \
b_{AT} = 1.9 {\rm \AA}^{-1}, \ b_{GC} = 2.5 {\rm \AA}^{-1}
$$ are given in \cite{CP} and used in \cite{ZC}.
The values
$$ D = 0.040 {\rm eV}, \ b = 4.45 {\rm \AA}^{-1} $$ are given in
\cite{PBD} and used in \cite{PBD,BCP,DeLuca}.
Finally,  the estimates
$$ D_{AT} = 0.050 {\rm eV}, \ D_{GC} = 0.075 {\rm eV}, \
b_{AT} = b_{GC} = 4 {\rm \AA}^{-1} $$ are given in \cite{Campa}
and used in \cite{Campa,KS}. The values of coupling constants
corresponding to these different values for the parameters
appearing in the Morse potential range across a whole order of
magnitude: \beq\lb{kd} 3.5 \, {\rm N/m} \ \leq \ K_p \, := \, 2
b^2 D \ \leq \ 38 {\rm N/m} \ . \feq

In our numerical investigations we will use a value of $K_{p}$
near to the lower bound given in \ref{kd}; that is, we adopt the
value $K_p = 4 {\rm N/m}$, leading to an optical frequency of
$\omega_0=\sqrt{2K_p/m}=36 {\rm cm}^{-1}$, so to be in agreement with
\cite{Pow}.
\bigskip

\subsubsection{Stacking}

The determination of the torsion and stacking coupling constants
is more involved  and rests on a smaller amount of experimental
data. The main information is the total torsional rigidity of the
DNA chain  $C=S \de$, where $\delta=3.4 {\rm \AA}$
is the base-pair spacing and $S$ is the torsional rigidity. It is
known \cite{BZ,BFLG} that \beq 10^{-28}\, {\rm J  m} \ \leq \ C \
\leq \ 4 \cdot 10^{-28} \, {\rm J  m} \ . \feq This information
is used e.g. in \cite{Eng,Zhang}, whose estimate is based on the
evaluation of the free energy of superhelical winding; this fixes
the range for the total torsional energy to be \beq 180 \, {\rm K
J/mol} \ \leq \ S \ \leq \ 720 \, {\rm K J/mol} \ . \feq

In our composite model the total torsional energy of the DNA chain
has to be considered as the sum of two parts, the base stacking
energy and the torsional energy of the sugar-phosphate backbone.
In order to extract the stacking coupling constant we use the
further information that $\pi-\pi$ stacking bonds amount at the
most to $50 {\rm KJ/mol}$ \cite{SC}. Assuming a quadratic stacking
potential, as we do, and a width of the potential well of about
$2\AA$  we obtain the estimate $K_s = 68 {\rm N/m}$. The phonon
speed induced by this is $c_1 = \de \sqrt{K_s/m} \simeq 6
{\rm Km/s}$, see eq. (\ref{speed}); this is rather close to the
the estimate of $1.8 {\rm Km/s} \leq c_1 \leq 3.5 {\rm Km/s}$
given in \cite{YakPRE}.

As we shall see in detail in Sect. \ref{s9}, choosing smaller
values for $K_s$ would have non-trivial consequences since
solitons with small topological numbers become unstable in the
discrete setting when the ratios $K_s/K_p$ and $K_t/K_p$ get small
enough (see sect.~\ref{s9}). In particular, this value for $K_s$
-- together with the $K_t$ below -- is barely enough to allow the
existence of solitons, as discussed later on this paper.

\subsubsection{Torsion and helicoidal couplings}
\label{sec:hel}

After extracting the stacking component, our estimate for the
torsional coupling constant $K_t$ is in the range \beq 130 \, {\rm
K J/mol} \ \leq \ K_t \ \leq \ 670 \, {\rm KJ/mol} \ . \feq Assuming
(see below) that $K_h \simeq K_t/25$, so that
$c_4=\sqrt{2K_t/I_s}$ (see eqs. (\ref{speed1}) and (\ref{speed})),
all of these values for $K_t$ induce phonon speeds slightly higher
with respect to the estimates cited above, between 5 Km/s and 11
Km/s. For our numerical investigations, to keep the phonon speed as
low as possible,  we will  set
$K_t=130 {\rm KJ/mol}$.

Finally, for the helicoidal coupling constant, following
\cite{GaeJBP}, we assume that $K_t$ and $K_h$ differ by about a
factor $25$, so that $K_h=5 {\rm KJ/mol}$.

\subsubsection{Discussion}

It is interesting to point out how the geometry of the model
nicely fits with the estimates of the binding energies so to
induce optical frequencies and phonon speeds of the right order of
magnitude (see also the discussion in Sect. \ref{s6}). This is not
the case in simpler models, where in order to get the right phonon
speed within a simple Y model one is obliged to assume for $K_t$
the unphysical value $K_t = 6000 {\rm KJ/mol}$ \cite{YakPRE}.

Our estimates, and hence our choices for the values of the
coupling constants appearing in our model, are summarized in
table~\ref{tab:dynParms}. We will use these values of the physical
parameters of DNA in the next sections, when discussing both the
linear approximation and the dispersion relations as well as the
full nonlinear regime and the solitonic solutions.

\begin{table}[ht]
\centering
   \begin{tabular}{|c|c|c|c|}
    \hline
    $K_t$ & $K_s$ & $K_p$ & $K_h$ \\
    \hline
    130 KJ/mol & 68 N/m & 3.5 N/m & 5 KJ/mol \\
    \hline
   \end{tabular}
   \caption{Values of the coupling constants for our DNA model}
   \label{tab:dynParms}
\end{table}

\section{Small amplitude excitations and dispersion relations}
\lb{s6} In this section we will investigate the dynamical behavior
of our model  for  small excitations in the linear regime. We will
enforce the Yakushevich condition $R + d_h = a$ in order to keep
the calculations and their results as simple as possible (see also
\cite{GaeY1}).

Linearizing the   equation  of motion (\ref{EuLag}) around the
equilibrium configuration
\beq \phi^{(a)}_i = \theta^{(a)}_i =
{\dot\phi}^{(a)}_i = {\dot\theta}^{(a)}_i =  0,  \feq
we get using standard algebra,
\beq\lb{linEL} \begin{array}{l}
m r^2 {\ddot \phi}^{(a)}_i + m (r R + r^2) {\ddot \theta}^{(a)}_i \ = \\
\ = \  K_s \[ (r R + r^2 ) (\theta^{(a)}_{i+1} - 2 \theta^{(a)}_i
+ \theta^{(a)}_{i-1}) +
        r^2 (\phi^{(a)}_{i+1} - 2 \phi^{(a)}_i + \phi^{(a)}_{i-1}) \] \ + \\
    \ \ \  - K_p \[
  (a-R)^2 (\phi^{(a)}_i + \phi^{(\hat  a)}_i) +
  a (a - R) (\theta^{(a)}_i + \theta^{(\hat  a)}_i ) \] \ ; \\
 {} \\
 m (r R + r^2 ) {\ddot \phi}^{(a)}_i + ( I + m (R+r)^2 ) {\ddot \theta}^{(a)}_i \ = \\
\ = \ K_t \[ \theta^{(a)}_{i+1} - 2 \theta^{(a)}_i +
\theta^{(a)}_{i-1} \] + \\
\ \ \ + K_s (r + R) \[ (R+r) (\theta^{(a)}_{i+1} - 2
\theta^{(a)}_i + \theta^{(a)}_{i-1}) +
        r (\phi^{(a)}_{i+1} - 2 \phi^{(a)}_i + \phi^{(a)}_{i-1}) \] + \\
        \ \ \ -  K_p \[ (a^2 -a R) (\phi^{(a)}_i + \phi^{(\hat  a)}_i) +
        a^2 (\theta^{(a)}_i + \theta^{(\hat  a)}_i ) \] + \\
        \ \ \ + K_h (\theta^{(\hat a)}_{i+5} - 2 \theta^{(a)}_i +
\theta^{(\hat a)}_{i-5}) \ .
\end{array} \feq
We are mainly interested in the  dispersion relations for the
propagating waves, which are solution of the system (\ref{linEL}).
To derive them it is convenient to introduce variables $\phi^{(\pm)}$ and
$\theta^{(\pm)}$ defined as
\beq\lb{tpm}
 \phi^\pm_i = \phi^{(1)}_i \pm \phi^{(2)}_i \ , \ \
\theta^\pm_i = \theta^{(1)}_i \pm \theta^{(2)}_i \ . \feq
Let us now  Fourier transform our variables, i.e. set
\beq\lb{fourier} \phi^\pm_n (t) \ = \ F^\pm_{k \om} \, \exp[i (k
\delta n + \om t) ] \ ; \ \theta^\pm_n (t) \ = \ G^\pm_{k \om} \,
\exp[i (k \delta n + \om t)] \ . \feq Here $k$ is the spatial wave
number, $\om$ is the wave frequency, and $\delta$ is a parameter
with dimension of length and set equal to the interpair distance
($\delta = 3.4~\AA$), introduced so that $k$ has dimension
$[L]^{-1}$ and the physical wavelength is $\la = 2 \pi / k$. In
this way, we should only consider $k \in [ - \pi / \delta , \pi /
\delta ]$.

Using (\ref{tpm}) and (\ref{fourier}) into (\ref{linEL}), we get a set of
linear equations for $(F^\pm_{k \om} , G^\pm_{k \om} )$; each set
of coefficients with indices $(k, \om )$ decouples from other wave
number and frequency coefficients, i.e. we have a set of four
dimensional systems depending on the two continuous parameters $k$
and $\om$. This is better rewritten in vector notation as
\beq\lb{4.6} M \, \zeta_{k \om} \ = \ 0, \feq where $\zeta_{k \om}$
is the vector of components \beq \zeta_{k \om} \ = \ \( F^+_{k
\om} \, , \, F^-_{k \om} \, , \,  G^+_{k\om} \, , \,  G^-_{k \om}
\) \feq and $M$ is a four by four matrix which we omit to write
explicitly.

In order to simplify the calculations we will set to zero the
radius of the disk modelling the base, i.e $d_{h}=r$. As in our
model the disk describing the base cannot rotate around its axis,
this assumption does not modify the physical outcome of the
calculations.
The condition for the existence of a solution to (\ref{4.6}) is
the vanishing of the determinant of $M$. By explicit computation
the latter is written as the product of three terms,
apart from a constant factor $r^2$,
\beq ||M|| \ = \ r^2
\, \pi_1 \, \pi_2 \, \pi_3 \ . \feq
The three factors being,
\bea\lb{p1}
\pi_1 \ &=& \ - 2 K_s + m \om^2 + 2 K_s
\cos ( \delta k) \ ,\nonumber\\
\pi_2 \ &=& \ I \om^2 - 2  ((K_h + K_t) -
K_t \cos[k \delta] +
       K_h \cos[5 k \delta]) \ ,\nonumber\\
\pi_3&=&  \left(I m r^2\right) \om^4 - 2  \ r^2 \,
\left( K_p I + 2 (K_s I + K_t m ) \sin^2 (k \de/2) +\right.\nonumber\\
&+&2 \left.m K_h \sin^2 (5 k \de /2) \right) \om^2 + \ 8 r^2 \, \(
K_p + 2 K_s \sin^2 (k \de / 2 ) \) \( K_h\sin^2 (5 k \de / 2) +
K_t \sin^2 (k \de /2 ) \). \eea The equation $||M||=0$ has four
solutions, given by \beq\label{drel}
\begin{array}{l}
\om^2_1 \ = \ 4 (K_s / m) \, \sin^2 (k \de/2 ) \ , \\
\om^2_2 \ = \ 4 (K_t / I) \, \sin^2 (k \de/2) \ + \ 2 (K_h /I )
\, [1+ \cos (5 k \de) ] \ , \\
\om^2_3 \ = \ 2 (K_p / m) \ + \ 4 (K_s/m) \, \sin^2 (k \de/2) \ ,
\\
\om^2_4 \ = \ 4 (K_t/I) \, \sin^2 (k \de/2) \ + \ 4 (K_h/I) \,
\sin^2 (5k \de /2) \ . \end{array} \feq

Eqs. (\ref{drel}) provide the dispersion relations for our model.
Physically, the four dispersion relations correspond to the four
oscillation modes of the system in the linear regime. The relation
involving $\om_{1}$ describes relative oscillations of the two
bases in the chain with respect to the neighboring bases. As
$\om_{1} (k)\to 0$ for $k\to 0$ there is no threshold for the
generation of these phonon mode excitations.

\begin{figure}
 \includegraphics[width=130pt]{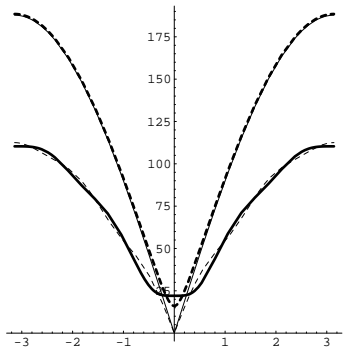}\\
 \caption{Graph of the dispersion relations (\ref{drel})
 in the first Brillouin zone.  We plot $\omega_{\alpha}/(2\pi
 c)$ ($c$  is the speed of light) as a function of $k\Delta/2$.
$\omega_1,\omega_2,\omega_3,\omega_4$ are represented respectively
by the thin continuous, thick continuous, thick dashed  and  thin
dashed line. Units are $cm^{-1}$ in the vertical axis and radiants
in the horizontal axis.}
 \label{fig:dispRel}
\end{figure}

The relations involving $\om_{2}$ and $\om_{4}$ are associated with
torsional oscillations of the backbone. In case of $\om_{2}$
there is a threshold for the generation of the excitation
originating in the helicoidal interaction, whereas the second
torsional mode $\om_{4}$ has no threshold and is thus also of
acoustical type.
The dispersion relation involving $\om_{3}$ describes relative
oscillations of two bases in a pair. The threshold for the
generation of the excitation is now determined by the pairing
interaction.
The dispersion relations (\ref{drel}) are plotted
as
$\om/(2\pi c)$, where $c$ is the speed of light  (we use the, in the
literature widespread, convention of measuring frequencies in $2\pi c$
units) versus $k\Delta/2$ in
Fig.~\ref{fig:dispRel} for values of the physical parameters
given in the tables  \ref{tab:cynParms}  and
\ref{tab:dynParms}.
The four dispersion relations take a simple  form if we consider
excitations with wavelength $\la$ much bigger then the intrapair
distance, i.e $\la >> \de$; this corresponds to the $\delta\to 0$
limit. We have then \beq \om^2_\a \, - \, c^2_\a \, k^2 \ = \
q^2_\a \ , \feq where $c_\a$ and $q_\a$ ($\a=1\ldots 4$) are, respectively, the
velocity of propagation (in the limit $k >> q_{\alpha}$) and the
excitation threshold. They are given by
\beq\lb{speed1} \begin{array}{ll}
c_1 = \delta \sqrt{K_s/m},\quad & q_1 = 0; \\
c_2 = \delta \sqrt{(K_t - 25 K_h )/I},\quad & q_2 = 2 \sqrt{K_h/I}; \\
c_3 = \delta \sqrt{K_s/m}, \quad & q_3 = \sqrt{2 K_p/m}; \\
c_4 = \delta \sqrt{(K_t + 25 K_h )/I},\quad & q_4 = 0. \end{array}
\feq
Using the values of the parameters given
in the tables \ref{tab:cynParms} and
\ref{tab:dynParms}
we have
\beq
\label{speed}
\begin{array}{ll}
  c_1 = 6.1 \, {\rm{ Km/s}},\quad & q_1 = 0 \ ; \\
  c_2 = 0 \, {\rm{ Km/s}},\quad & q_2 = 22 \, {\rm{ cm}^{-1}} \ ; \\
  c_3 = 6.1 \, {\rm{ Km/s}},\quad & q_3 = 36 \, {\rm {cm}}^{-1} \ ; \\
  c_4 = 5.1 \, {\rm {Km/s}},\quad & q_4 = 0 \ ,
\end{array}
\feq
 where $c_2=0$ comes from the fact that we are taking $K_t\simeq25K_h$
(see  table \ref{tab:dynParms}). This of course  just means that
$c_2$ is at least an order of magnitude smaller than the other
$c_i$ -- and therefore negligible. Speeds can be converted to base
per seconds by dividing each $c_i$ by $\delta=3.4\AA$; excitation
thresholds can be converted in inverse of seconds by multiplying
each $q_i$ by $2\pi c$, where $c$ is the speed of light.

\section{Nonlinear dynamics and travelling waves}
\lb{s7}
After studying the small amplitude dynamics of our model, we
should now investigate the fully nonlinear dynamics. We are in
particular interested in soliton solutions, and on physical basis
they should have -- if the model has any relation with real DNA --
a size of about twenty base pairs. This also means that such
solutions vary smoothly on the length scale of the discrete chain,
and we can pass to the continuum approximation. On the other hand,
such a smooth variance assumption is not justified on the length
scales (five base pairs) involved in the helicoidal interaction,
and one should introduce nonlocal operators in order to take into
account helicoidal interactions in the continuum approximation
\cite{GRPD}.

Luckily, numerical experiments show that -- at least in the case
of the original Yakushevich model -- soliton solutions are very
little affected by the presence or  of the helicoidal
terms (as could also be expected by their intrinsical smallness,
in a context where they cannot play a qualitative role as for
small amplitude dynamics) \cite{GRPD}. Thus, we will from now on
simply drop the helicoidal terms, i.e. set $K_h = 0$.

\subsection{Continuum approximation and field equations}

The continuum description of the discrete model we are considering
requires to introduce fields $\Phi^{(a)} (z,t) , \Theta^{(a)}
(z,t) $ such that \beq\lb{6.1} \Phi^{(a)} (n \de ,t) \approx
\phi^{(a)}_n \ , \ \Theta^{(a)} (n \de ,t) \approx \theta^{(a)}_n
\ . \feq

The continuum approximation we wish to consider is the one where
we take \beq\lb{6.2} \begin{array}{l} \Phi (x \pm \de , t) \
\approx \ \Phi (x , t) \ \pm \ \de \,
\Phi_x (x,t) \ + \ (1/2) \, \de^2 \, \Phi_{xx} (x,t) \ , \\
\Theta (x \pm \de , t) \ \approx \ \Theta (x , t) \ \pm \ \de \,
\Theta_x (x,t) \ + \ (1/2) \, \de^2 \, \Theta_{xx} (x,t) \ .
\end{array} \feq

Inserting (\ref{6.2}), and taking $K_h=0$, into the Euler-Lagrange
equations (\ref{EuLag}), we obtain a set of nonlinear coupled PDEs for
$\Phi^{(a)} (x,t)$ and $\Theta^{(a)} (x,t)$, depending on the
parameter $\de$. Coherently with (\ref{6.2}), we expand these equations to
second order in $\de$ and drop higher order terms. The equations
we obtain in this way  are symmetric in the chain exchange.
We will be able to decompose their solutions into a symmetric and
an antisymmetric part under the same exchange. In view of the
considerable complication of the system, it is convenient to deal
directly with the equations for these symmetric and antisymmetric
part and to enforce the Y contact approximation $R+d_{h}=a$

That is  we will consider fields \beq\lb{6.4} \Phi^\pm =
\Phi^{(1)} \pm \Phi^{(2)} \ , \ \Theta^\pm = \Theta^{(1)} \pm
\Theta^{(2)} \ , \feq and discuss equations which result by
setting two of these to zero. In the symmetric case, i.e. for
\beq\lb{6.5} \Theta^{(1)} (x,t) = \Theta^{(2)} (x,t) = \Theta
(x,t) \ , \ \Phi^{(1)} (x,t) = \Phi^{(2)} (x,t) = \Phi (x,t) \ ,
\feq the resulting equations are
\beq\lb{6.6} \begin{array}{l} m
r^2 \Phi_{tt} + (mr^2 + m R r \cos \Phi ) \Theta_{tt} \ = \ - 2
a (a-R) K_p \sin (\Phi + \Theta ) + \\
\ \ \ - R (2 K_p (R-a) + m r
\Theta_t^2 ) \sin \Phi + \\
\ \ \ + \de^2 \ K_s r \[ r (\Phi_{xx} + \Theta_{xx} ) + R
\Theta_{xx} \cos \Phi + R \Theta_x^2 \sin \Phi \] \ ; \\
( m r^2 + m R r \cos \Phi ) \Phi_{tt} + (I + m (R^2 + r^2 + 2 R r
\cos \Phi ) ) \Theta_{tt} \ = \\
\ \ \ = - 2 a K_p (R \sin \Theta + (a - R) \sin (\Phi + \Theta) )
+ m \Phi_t R r (\Phi_t + 2 \Theta_t ) \sin \Phi + \\
\ \ \ + \de^2 \[ K_s r^2 \Phi_{xx} + (K_t + K_s (R^2 + r^2 ))
\Theta_{xx} + \right. \\
\ \ \ \left. + K_s R r \( (\Phi_{xx} + 2 \Theta_{xx}) \cos \Phi -
\Phi_x (\Phi_x + 2 \Theta_x ) \sin \Phi \) \] \ .
\end{array} \feq
In the antisymmetric case
\beq\lb{6.7} \Theta^{(1)} (x,t) = -
\Theta^{(2)} (x,t) = \Theta (x,t) \ , \ \Phi^{(1)} (x,t) = -
\Phi^{(2)} (x,t) = \Phi (x,t) \ , \feq the resulting equations are
\beq\lb{6.8} \begin{array}{l}
m r^2 \Phi_{tt} + (m r^2 + m R r \cos \Phi ) \Theta_{tt} \ = \\
\ \ \ = \ K_p (a-R) \( R \sin (\Phi + 2 \Theta ) + (a - R) \sin (2
(\Phi + \Theta ) ) - 2 a \sin (\Phi + \Theta ) \) \ + \\
\ \ \ + R (K_p (a-R) - m r \Theta_t^2 ) \sin \Phi + \\
\ \ \ + \de^2 K-s r \[ r (\Phi_{xx} + \Theta_{xx} ) + R \cos
(\Phi) \Theta_{xx} + R \sin (\Phi ) \Theta_x^2 \] \ ; \\
(m r^2 + m R r \cos \Phi ) \Phi_{tt} + (I + m (R^2 + r^2 + 2 R r
\cos \Phi)) \Theta_{tt} \ = \\
\ \ \ = \ K_p \( -2 a R \sin \Theta + R^2  \sin (2 \Theta) + \right. \\
\ \ \ \left. + (a - R) (-2 a \sin (\Phi + \Theta) + (a - R) \sin
(2 (\Phi + \Theta)) +
2 R \sin (\Phi + 2 \Theta) ) \) + \\
\ \ \ + m \Phi_t r R (\Phi_t + 2 \Theta_t) \sin \Phi + \\
\ \ \ + \de^2 \[ K_s r^2 \Phi_{xx} + K_t \Theta_{xx} + K_s (R^2 +
r^2 ) \Theta_{xx} + \right. \\
\ \ \ \left. + K_s r R ((\Phi_{xx} + 2 \Theta_{xx}) \cos \Phi -
\Phi_x (\Phi_x + 2 \Theta_x) \sin \Phi)\] \ . \end{array} \feq We
will not write the equations in the cases of mixed symmetry, i.e.
for $\Theta^{(1)} (x,t) = \Theta^{(2)} (x,t) = \Theta (x,t)$,
$\Phi^{(1)} (x,t) = - \Phi^{(2)} (x,t) = \Phi (x,t)$ and for
$\Theta^{(1)} (x,t) = - \Theta^{(2)} (x,t) = \Theta (x,t)$ and
$\Phi^{(1)} (x,t) = \Phi^{(2)} (x,t) = \Phi (x,t)$.

\subsection{Soliton equations}

When studying DNA models, one is specially interested in
travelling wave solutions, i.e. solutions depending only on $z :=
x - v t$ with fixed speed $v$: \beq\lb{6.9} \Phi^{(a)} (x,t) =
\vphi^{(a)} (x - v t) := \vphi^{(a)} (z) \ , \ \Theta^{(a)} (x,t)
= \vth^{(a)} (x-vt) := \vth^{(a)} (z) \ . \feq

If we insert the ansatz (\ref{6.9}) into the equations (\ref{6.6}) and
(\ref{6.8}), we get a set
of four coupled second order ODEs; defining
\beq\lb{6.10} \mu \ :=
\ (m v^2 - K_s \de^2 ) \ , \ \ J \ := (I v^2 - K_t \de^2 ) \ ,
\feq
in the completely symmetric case (\ref{6.6}) we obtain

\beq\lb{6.12}
\begin{array}{l}
\mu r^2 \, \vphi'' \ + \ \mu r (r + R \cos \vphi ) \, \vth'' \ = \\
\ \ = \ - 2 a K_p (a - R) \, \sin \(\vphi + \vth \) + K_s \de^2 R
r \, \sin \( \vphi \) \, (\vth' )^2 \ + \\
\ \ \ - R \, \sin (\vphi) \, (-2 K_p (a - R) + m r v^2 (\vth' )^2
) \ ; \\
\mu r (r + R \cos \vphi ) \, \vphi'' \ + \ [ J + \mu  (R^2 + r^2 +
2 R r \cos \vphi ) \, \vth'' \ = \\
\ \ = \ - 2 a K_p \( R \sin \vth + (a - R) \sin (\vphi + \vth) \)
+ \mu R r \, \sin (\vphi) [ (\vphi')^2 + 2 \vphi' \vth' ] \ .
\end{array} \feq

In the completely antisymmetric case (\ref{6.8}), instead, we get
\beq\lb{6.13}
\begin{array}{l}
\mu r^2 \, \vphi'' \ + \ \mu r (r + R \cos \vphi ) \, \vth'' \ = \\
\ \ = \ - 2 K_p (a - R) (a - R \cos \vth - (a - R) \cos (\vphi +
\vth )) \sin (\vphi + \vth ) + \\
\ \ \ - \mu  R r (\sin
\phi) (\vth' )^2 \ ; \\
\mu r (r + R \cos \vphi ) \, \vphi'' \ + \ [ J + \mu (R^2 + r^2 +
2 R r \cos \vphi )
\, \vth'' \ = \\
\ \ = \ - K_p \[ 2 a R \sin \vth - R^2 \sin (2 \vth ) + \right.
\\
\ \ \ \left. + (a -  R) \(2 a \sin (\vphi + \vth) - (a - R) \sin
(2 (\vphi + \vth)) - 2 R
\sin (\vphi + 2 \vth) \) \] + \\
\ \ \ + \mu r R (\sin \vphi) [(\vphi' )^2  + 2 \vphi' \vth' ] \ .
\end{array} \feq

The previous equations appear too involved to be studied analytically
at least in the general case.  Numerical
results are discussed in Sect. \ref{s9}  below.
Some understanding can be gained  at the analytical level by considering
a particular case of the full equations (\ref{6.12}),(\ref{6.13}), when
the system reduces essentially to the Y case.
The next section is devoted to this.

\subsection{Boundary conditions}

We have so far just discussed the field equations (\ref{6.6}) and
(\ref{6.8}) and their reductions; however these PDEs make sense
only once we specify the function space to which their solutions
are required to belong.
The natural physical condition is that of {\it finite energy}; we
now briefly discuss what it means in terms of our equations and
the boundary conditions it imposes on their solutions.

The field equations (\ref{6.6}), (\ref{6.8}) are Euler-Lagrange
equations for the Lagrangian obtained as continuum limit of
(\ref{lag}). In the present case, the finite energy condition
corresponds to requiring that for large $|z|$ the kinetic energy
vanishes and the configuration correspond to points of minimum for
the potential energy.
The condition on kinetic energy yields \beq\lb{finkin} \Phi_t (\pm
\infty,t) = 0 \ , \ \Theta_t (\pm \infty,t) = 0 \ , \feq where of
course $\Phi_t (\pm \infty,t)$ stands for $\lim_{z \to \pm \infty}
\Phi_t (z,t)$, and so for $\Theta_t$.

As for the condition involving potentials, by the explicit
expression of our potentials, see above, this means (with the same
shorthand notation as above) \beq\label{finpot}
\begin{array}{l} \Phi (\pm \infty,t) = 0 \ , \ \Theta (\pm \infty,t) =
2 n_\pm \pi \ , \\
\Phi_z (\pm \infty,t) = 0 \ , \ \Theta_z (\pm \infty,t) = 0 \ ;
\end{array} \eeq

Let us now consider the reduction to travelling waves, i.e. eqs.
(\ref{6.12}) and (\ref{6.13}). In this framework, conditions
(\ref{finkin}) and (\ref{finpot}) imply we have to require the
limit behavior described by \beq\label{fes}
\begin{array}{l} \vphi
(\pm \infty) = 0 \ , \ \vth (\pm \infty) = 2 n_\pm \pi \ , \\
\vphi' (\pm \infty) = 0 \ , \ \vth' (\pm \infty) = 0 \ ,
\end{array} \eeq for the functions $\vphi (\xi)$, $\vth (\xi )$.

We would like to stress that eqs. (\ref{6.12}) and (\ref{6.13})
can also be seen as describing the motion (in the fictitious time
$\xi$) of point masses of coordinates $\vth (\xi), \vphi (\xi)$ in
an effective potential; such a motion can satisfy the boundary
conditions (\ref{fes}) only if $(\vphi,\vth) = (0,2 \pi k )$ is a
point of maximum for the effective potential. This would provide
the condition $\mu < 0$ and hence a maximal speed for soliton
propagation (as also happens for the standard Y model); we will
not discuss this point here, as it is no variation with the
standard Y case, and the condition $\mu < 0$ is satisfied with the
values of parameters obtained and discussed in Sect. \ref{s5}.

The solutions satisfying (\ref{fes}) can be classified by the
winding number $n := n_+ - n_-$.
Needless to say, here we considered the equations describing
symmetric or antisymmetric solutions, but a similar discussion
also applies to the full equations, i.e. those in which we have
not selected any symmetry of the solutions; in this case we would
have two winding numbers, which we can associated either to
$\vth^{(1)}$, $\vth^{(2)}$ or directly to their symmetric and
antisymmetric combinations $\vth^{(1)} \pm \vth^{(2)}$.

\section{Comparison with the Yakushevich model}
\lb{s8}

The standard Y model \cite{YakPLA} can be seen as a particular
limiting case of our model. Thus, a  check of the validity of our
model -- and in particular of the fact we considered here physical
assumptions which correspond to those by Y in her geometry -- can
be obtained by going to a limit in which our composite Y model
reduces to the standard Y model.
The latter can be obtained as a limiting case of our composite
model in two conceptually different ways.
\begin{itemize}
\item A first possibility, which we call {\it parametric}, is to
choose the geometrical parameters of the model so that its
geometry actually reduces to that of the standard Y model. \item A
second possibility, which we call {\it dynamical} is to force the
dynamics of our model by setting $\varphi^{a}=0$, i.e. by freezing
the non-topological angles and constraining them to be zero.
\end{itemize}

Let us briefly discuss these in some more detail. The parametric
way to recover the standard Y model from our model consists in
setting to zero the radius of the disks modelling the bases, and
at the same time pushing it on the disk representing the
nucleoside on the DNA chain. In this way the base corresponds to a
point on the circle bounding the disk representing the nucleoside.
Note that this would cause a change in the interbase equilibrium
distance, unless we at the same time also change the radius of the
disk representing the nucleoside.

This limiting procedure corresponds -- recalling we also want to
recover the Y approximation of zero interbase distance -- to the
following choice of the parameters:
\beq\lb{8.1}
\ m \ = \ K_{s}=d_{h} \ = \ r \ = \ 0 \ , \ \ a \ = R \ .
\feq

After the base has been pushed on the disk, its mass enters to be
part of the disk's mass -- hence contributes to its moment of
inertia -- and we can thus just take $m=0$. Similarly, as the
bases have lost their identity and are enclosed in the disk
modelling the whole nucleotide, the effective stacking interaction
has to be physically identified with the torsional interaction of
the disk now modelling the entire nucleotide. For this reason in
our equations we will take $K_s=0$ and $K_t \to K_s$.

Use of equations (\ref{8.1}) into the equations of motion (\ref{EuLag})  yields
\beq\lb{8.2}
\begin{array}{rl}
I {\ddot \theta}^{(a)}_{i} \ =& \ K_s  \sin ( \theta^{(a)}_{i-1} -
\theta^{(a)}_{i} ) \, + \, K_s \sin (\theta^{(a)}_{i+1} - \theta^{(a)}_{i} ) \,
+ \\
& \ + \, K_p R^2 \sin ( \theta^{(a)}_{i} - \theta^{(\hat a)}_{i})
   \, + \, K_h \( \theta^{(\hat a)}_{i+5} - 2  \theta^{(a)}_{i} +
   \theta^{(\hat a)}_{i-5} \).
\end{array}
\feq The previous equations represents the equations of motions
for the Y model. Some care has to be used when the values of the
parameters given by Eq. (\ref{8.1}) correspond to singular points
of the equations. This is for instance the case of the dispersion
relations $\omega_{1}$,$\omega_{3}$  in Eqs. (\ref{drel}),  which
are singular for $m=0$. The dispersion relations for the Y model
can be easily found by linearizing the system (\ref{8.2}). One
finds two dispersion relations; one is given by  $\om_2$ of the
composite model, see Eqs. (\ref{drel}); the other is \beq\lb{8.3}
\om^{2} \ = \ 2 R^{2}\frac{K_{p}}{I} \ + \ \frac{4
K_{s}}{I}\sin^{2} (\frac{k \de}{2}) \ + \ \frac{4
K_{h}}{I}\sin^{2} (\frac{5k \de}{2}) \ . \feq

To obtain the Y model dynamically from our model, we set $\Phi=0$
into the continuum equations (\ref{6.6}) and (\ref{6.8}). We also
enforce the Y condition $R+d_{h}=a$ and work in the zero
radius approximation for the disk modelling the base, i.e we set
$r=d_{h}$. In the fully symmetric case we get from Eqs.
(\ref{6.6})
\beq\lb{8.4}
\begin{array}{l}
m \Theta_{tt} \ = \ - 2 K_p \sin ( \Theta ) \, + \, \de^2 K_s \Theta_{xx} \ \ ;
\quad \(\frac{I}{(R + r)^2} + m \) \, \Theta_{tt} \
 = \ - 2K_{p} \sin \Theta \, + \, \de^2 \[ \frac{K_t}{(r+R)^2 } + K_s  \]
\Theta_{xx}. \  \end{array}
\feq
Compatibility of the previous
two equations requires that
\beq\lb{8.5}
I \, \Theta_{tt} \ = \ \de^{2} \, K_{t} \ \Theta_{xx} \ .
\feq
In the case of travelling wave solutions $\Theta(x,t)=\vth(x-vt)$
the constraint (\ref{8.5}) reads
\beq\lb{8.6}
 v^{2} \ = \ \frac{\de^2}{I} \, K_{t}.
\feq
We take from now on the
positive determination of velocity for ease of discussion.
Using the  Eqs. (\ref{6.9}),  (\ref{6.10}) and  (\ref{8.6}),
Eqs. (\ref{8.4})  yields
the travelling wave equation,
\beq\lb{6.20} \vth'' \ = - 2 (K_p / \mu_0 ) \ \sin \vth \,
\feq
where
\beq\lb{6.19}  \mu_0= \ (m K_t - I K_s ) \ \frac{\de^2} { I} \ . \feq

With the usual boundary conditions $\vth'(\pm\infty)=0$,
$\vth(\infty)= 2\pi$, $\vth (- \infty ) = 0$, Eq. (\ref{6.20}) has a
solution for $\mu_0 <0$, given precisely by the  $(1,0)$ \Y
soliton \beq\lb{6.21} \vth \ = \ 4 \ {\rm arctan} \[ e^{4\kappa z}
\] \ , \ \ \ \kappa = \sqrt{2K_{p}|\mu_{0}|} \ . \feq
We have thus recovered for the topological angles -- imposing the
vanishing of non-topological angles as an external constraint --
the Y solitons.
The condition for the existence of the soliton, $\mu_0<0$,
 implies that the physical parameters of
our model must satisfy the condition
\beq\lb{6.21a} \frac{K_{t}}{I}
\ < \ \frac{K_{s}}{m} \ . \feq
With the parameter values given in
the tables  \ref{tab:cynParms} and \ref{tab:dynParms}, we have
\beq\lb{6.22} \frac{m K_t} {I K_s } \ \simeq \ 0.3
\ < \ 1 \ ; \feq hence (\ref{6.21a}) is satisfied and we are in the
region of existence of the soliton.

Let us now consider  the antisymmetric equations.
Using $\phi=0$ into the system (\ref{6.13}) and the compatibility
equation (\ref{8.6}) we get
\beq\lb{6.23} \vth'' \ = - 2 (K_p
/\mu_0) \ (1 - \cos \vth) \ \sin \vth \ . \feq
As expected this
equation, with the usual boundary conditions (see above), admits a
solution for $\mu_0<0$, and the solution is in this case given by
$(0,1)$ \Y soliton \beq\lb{6.24} \vth \ = \ - 2 {\rm arccot} \[ -
\kappa z \] \feq (with $\kappa$ as above). The allowed values of
the physical parameters are determined by the same arguments used
for the $(1,0)$ soliton.

It should be stressed that in the standard Y model the travelling
waves speed is essentially a free parameter, provided the speed is
lower than a limiting speed \cite{GaeSpeed,GaeJBP}. Here,
recovering the standard Y model as a limiting case of the
composite model produces a selection of the soliton speed given by
Eq. (\ref{8.6}); this makes quite sense physically, as it
coincides with the speed of long waves determined by the
dispersion relations (\ref{speed1}).

\section{Numerical analysis of soliton equations and soliton solutions}
\label{s9}

Even after the several simplifying assumptions we made
for our DNA model, the complete equations of motions
given by Eqs. (\ref{6.6}) and (\ref{6.8}) respectively for
symmetric and antisymmetric configurations,  are too complex to be
solved analytically; the same applies to the reduced equations
(\ref{6.12}), (\ref{6.13}) describing soliton
solutions. We will thus look for solutions, and in particular for
the soliton solutions we are interested in, numerically.

In order to determine the profile of the soliton solutions we will
analyze the stationary case, with zero speed and kinetic energy,
and apply the ``conjugate gradients'' algorithm (see e.g.
\cite{NR,gsl}) to evaluate numerically the minima of our
Hamiltonian

This approach also allows a direct comparison with the results
obtained for the standard \Y model, and shown in \cite{YakPRE},
where authors proceed in the same way and by means of the same
numerical algorithm; this again with the aim, as remarked several
times above, to emphasize the differences which are due purely to
the different geometry of our ``composite'' model. With the same
motivation, we have also checked our numerical routines by
applying them to the standard Y model; in doing this we have also
considered with some care -- and fully confirmed -- certain
nontrivial effects mentioned in \cite{YakPRE}.

\subsection{Solitons in the standard \Y model}

As mentioned above, we will first present   the  numerical
investigation of  the stationary solitons of \Y model. Although
the soliton solutions of the \Y model have already been the
subject of previous numerical investigations \cite{YakPRE}, it is
useful to repeat the analysis here in order to check our algorithm
(and confirm the results reported by \cite{YakPRE}).

Moreover, in the next section we will compare the soliton
solutions of our composite model with those of the \Y model. We
will therefore need an explicit numerical results for the \Y model
obtained using our code.

The homogeneous \Y Hamiltonian for static solutions, i.e. setting
the kinetic term $T$ to zero, is \beq\label{eq:YakHam}
  \begin{array}{rl}
    H_Y \ =& \ \frac{I}{2} \, \sum_{i,a} \, (\Dpsi)^2 +
    g \, \sum_{i,a} \, \sin^2 ((\Dpsi)/2) + K \sum_i \[3 + \cos (\theta^1_i-\theta^2_i) -
    2 \cos\theta^1_i - 2 \cos\theta^2_i \]\\
    =& \ I \, \sum_{n} \, [\dpsip^2 + \dpsim^2] +
    g \, \sum_n \, [1-\cos\dpsip\cos\dpsim]
    + 2 K \sum_n \[1 - 2\cos\psip_n\cos\psim_n + \cos^2\psim_n \] \ .
  \end{array}
\feq
Here $I$ is the inertia momentum; $K$ and $g$ respectively
the pairing and torsional coupling constants of the \Y model;
$\theta^{(1,2)}$ are the angles describing the sugar rotation with
respect to the backbone. Moreover, we write $\psip=\theta^+$,
$\psim=\theta^-$ (the $\theta^{\pm}$ are defined as in
Eq. (\ref{tpm})), and $\Delta_n\psip :=\psip_{n+1}-\psip_n$ and
similarly for $\psim$.

If we select the physical value for $I$ (given in table
\ref{tab:cynParms}) and factor out the $K$
(equivalently, we measure energies in units of $K$; this takes the
value $K=150$KJ/mol), then the only independent parameter left in
$H$ is the coupling constant $g$. This can and will be used,
as in \cite{YakPRE}, for a parametric study of the solutions.
In our numerical investigations we used (in order to avoid any
accidental spurious effect) two independent implementations of the
conjugate-gradient
algorithms, i.e. the one
developed by Numerical Recipes \cite{NR} and the one provided by
the GNU \cite{gsl}. The results obtained with the two
implementations turned out to be in very good agreement.

The conjugate-gradient algorithms requires a ``starting point'',
 i.e. an initial approximation of the minimizing
configuration; in the case of multiple minima, the algorithm will
actually determine a local minimum or the other depending on this
initial approximation. Following \cite{YakPRE}, we have used as
starting points hyperbolic tangent profiles
\beq\lb{start} \theta^{1,2}_n \ =
\ q^{1,2} \ \pi (1 + \tanh (\beta (2n-N))) \ , \feq
where
$(q^1,q^2)$ is the topological type of the soliton, $N$ is the
number of sites in the chain, and $\beta$ a parameter, whose
reasonable range is about $0 \leq \beta \leq N/10$, used to adjust
the profile.
The parameter $\beta$ is crucial to determine the structure of the
minima (hence of the solitonic solutions) of the \Y Hamiltonian.
Choosing  different ranges of variation for $\beta$ we are probing
different dynamical regions of the system, where different local
minima of the Hamiltonian  may be present.

In the case of the ``elementary'' solitons -- the $(1,0)$ and the
$(0,1)$ ones -- for which only one degree of freedom matters, we
obtain the same local minimum, up to $10^{-5}$ in the energy and
$10^{-2}$ in the angles, independently of $\beta$ (provided of
course that $\beta$ is not too close to zero; in our case it
suffices to keep $\beta \geq 4$) in agreement with the fact that
in this case the minimum is known to be global.

\subsubsection {Quasi-degeneration of the energy minimizing configurations
for higher topological numbers}

The situation appears to be different in the case of the $(1,1)$
soliton. Here in facts numerical investigations
show a strong dependence on $\beta$ of the local minimum
determined by the algorithm for most of its range, in particular
for $\beta > 6$,  while energies vary very little -- within
$0.2\%$ -- about 49.3K. This  suggests we are in the presence of a rather
complex structure of the phase space for the (1,1) limiting
conditions. There still is however a small interval, ($4 \leq
\beta \leq 6$), where the algorithm behaves exactly as in the
$(1,0)$ and $(0,1)$ cases, i.e. yields almost the same energy minimizing
configuration as $\beta$ is varied, and allows us to find what we
believe is the global minimum of the Hamiltonian.

We have detected this same behavior also in solitons of higher
topological types, e.g. $(2,0)$ and $(2,1)$. This suggests we are
in the presence of a generic pattern \footnote{There are indeed
qualitative arguments suggesting a degeneration of minimizing
configuration whenever the topological numbers are not (1,0) or
(0,1); we will not discuss these arguments nor the matter here.};
we point out that the reason why this same behavior is not shared
by the solitons of types $(1,0)$ and $(0,1)$ is related to the
fact that in these two cases (and only in them) the problem
reduces actually to a one-dimensional dynamics, while all others
cases are intrinsically two-dimensional.

\begin{figure}
$$\begin{array}{cc}
    \includegraphics[width=200pt]{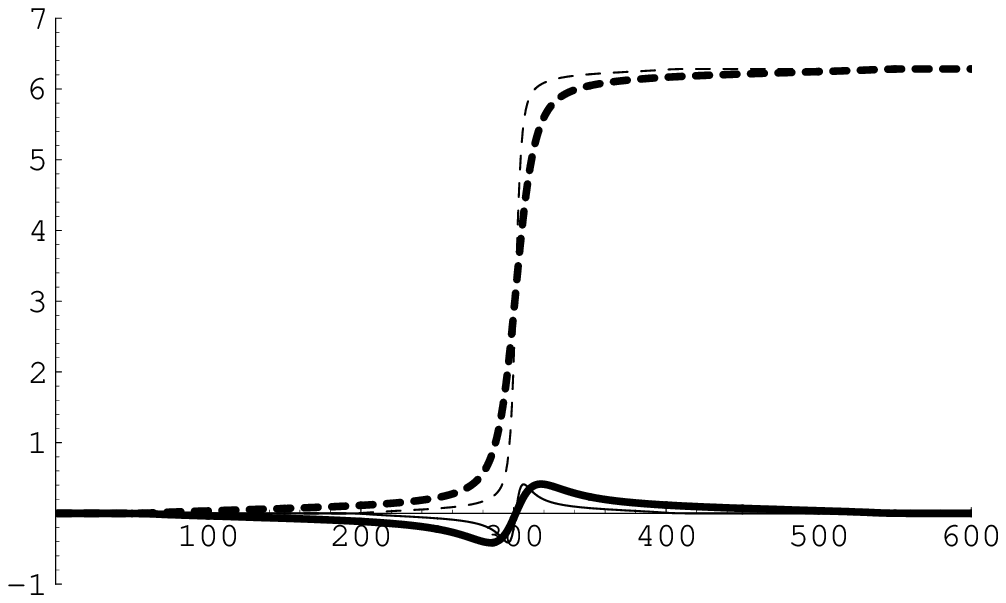} &
    \includegraphics[width=200pt]{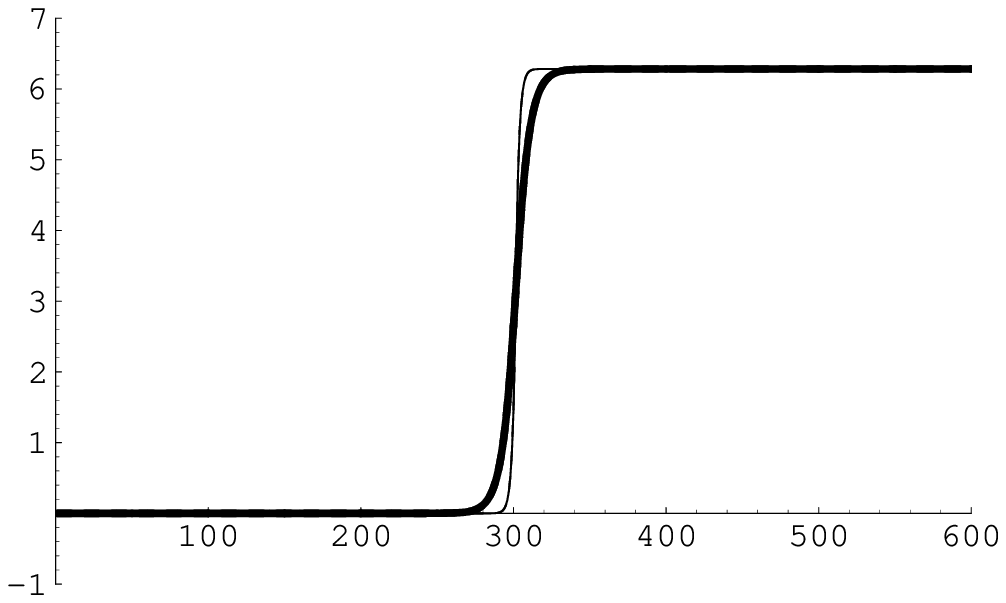}\\
    a & c \\
    \includegraphics[width=200pt]{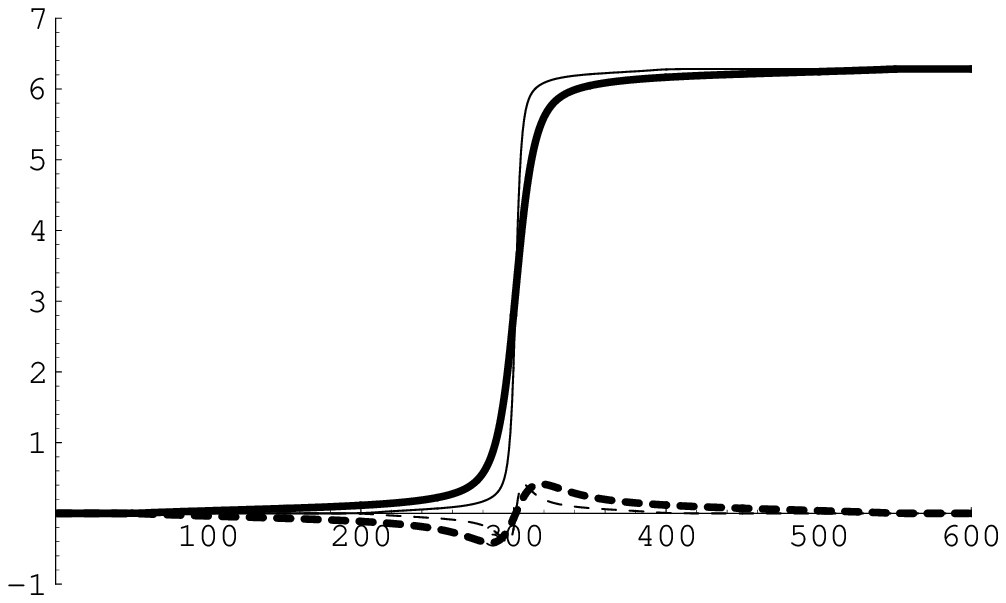} &
    \includegraphics[width=200pt]{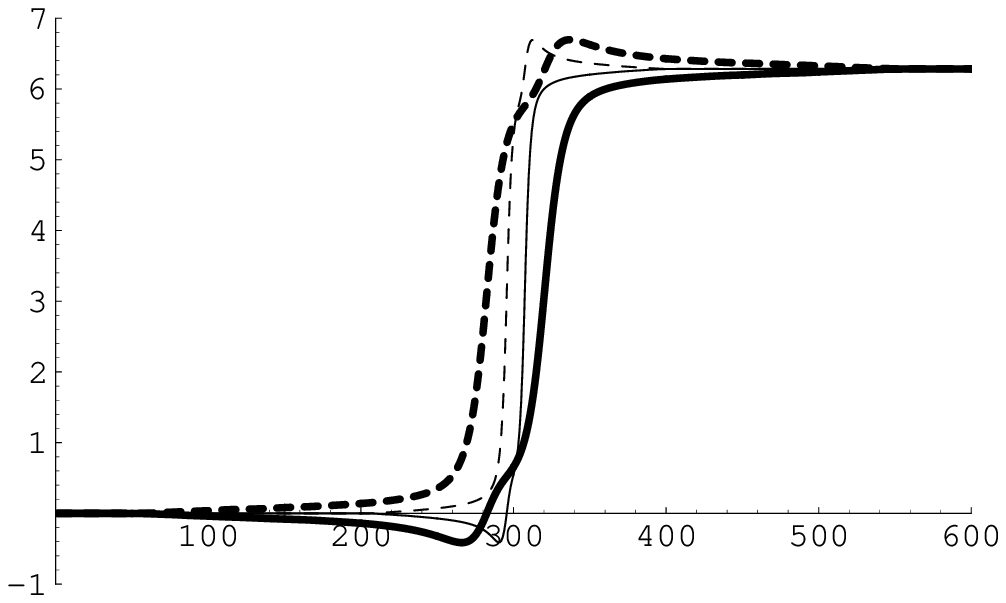} \\
    b & d \end{array}$$
  \caption{Static solitons profiles for the Yakushevich homogeneous Hamiltonian.
  Thicker lines correspond to $g=150$, thinner ones to $g=23.4$;
  dashed ones to the angle
  $\theta^1$ and continuous ones to the angle $\theta^2$. Topological
  numbers are as follows. Upper left (a): (1,0); lower left (b):
  (0,1); upper right (c): (1,1); lower right (d): (1,1). Picture
  (d) is obtained by a computation where $2\pi$ has been approximated with 6.28,
  and shows a sensitive dependence of the solution on the boundary conditions.
  The energies $E^g_{(p,q)}$ of the solitons are:
  \energy{150}{1}{1}{153.4\, K}, \energy{150}{0}{1}{62.93\, K},
  \energy{150}{1}{0}{62.93\, K}, \energy{23.4}{1}{1}{59.32\, K},
  \energy{23.4}{0}{1}{24.63\, K}, \energy{23.4}{1}{0}{24.63\, K}.
  Energies are measured in units of $K=150 {\rm KJ/mol}$.} \label{Yak}

\end{figure}

\begin{figure}
$$\begin{array}{ccc}
    \includegraphics[width=150pt]{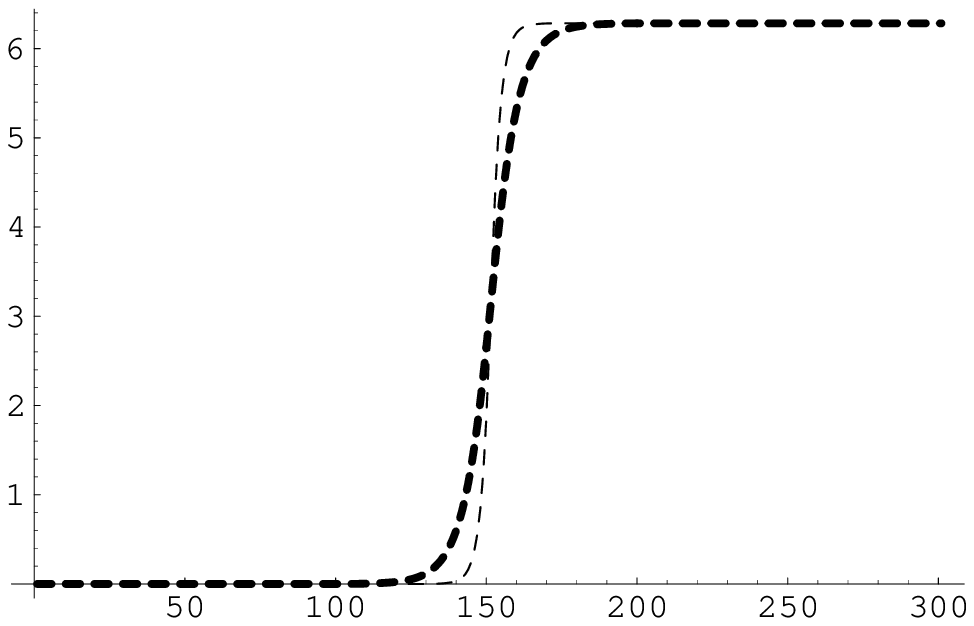}&
    \includegraphics[width=150pt]{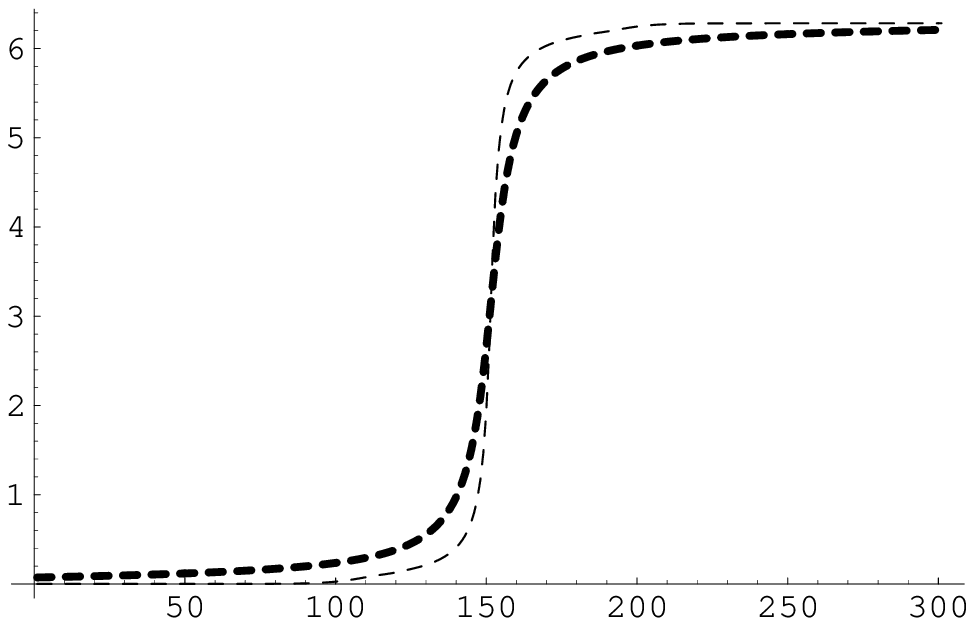}&
    \includegraphics[width=150pt]{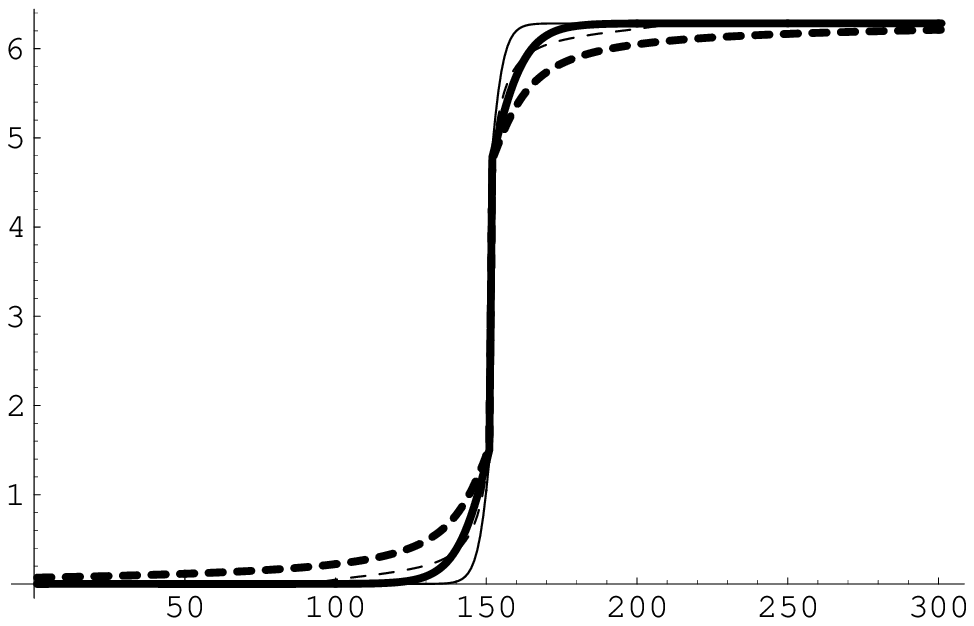}\\
    a&b&c\end{array}$$
    \caption{Static solitons profiles for the Yakushevich homogeneous
      hamiltonian expressed in the coordinates $(\psip,\psim)$.
      Thicker lines correspond to $g=150$, thinner ones to $g=23.4$;
      dashed ones to the angle $\psip$ and continuous ones to
      the angle $\psim$. Topological numbers are: (1,0) in
      (a), (0,1) in (b), (1,1) in (c).
      The energies $E^g_{(p,q)}$ of the solitons are:
      \energy{150}{1}{1}{62.93\,K}, \energy{150}{0}{1}{153.4\,K},
      \energy{150}{1}{0}{195.3\,K}, \energy{23.4}{1}{1}{24.64\,K},
      \energy{23.4}{0}{1}{59.34\,K}, \energy{23.4}{1}{0}{75.56\,K}.
      Energies are measured in units of $K=150 {\rm KJ/mol}$.}
     \label{Yak2}
\end{figure}

\subsubsection{Numerical instability of the solitons}

A noteworthy fact pointed out in \cite{YakPRE} is that the
discrete version of the soliton solutions lose stability,
namely switch from minima to saddle points,
as $g$ gets close to $0$, a phenomenon that is not shared
by its continuos counterpart.
We have looked for this effect and confirmed the findings of
\cite{YakPRE}. In our numerical computations the $g$ values at
which the transition  takes place turned out to be different
if computations are performed in the $\theta^{1,2}$ coordinates or
in the $(\psi,\chi)$ ones. Transition values corresponding to the
onset of the numerical instability are given in Table \ref{tab:instab}

\begin{table}
  \centering
  \begin{tabular}{|c||c|c|c|}
  \hline
    & (1,0) & (0,1) & (1,1) \\
  \hline
  ($\theta^1,\theta^2$) & 7.05 & 7.05 & 14.7 \\
  ($\psi,\chi$) & 14.7 & 16.2 & 7.0 \\
  \hline
\end{tabular}
  \caption{The transition values of $g_0$ for instability (arising
  for $g < g_0$)
  of the $(p,q)$ solitons.}\label{tab:instab}
\end{table}

The transition values are, in the $(\theta^{1,2})$ coordinates,
$g=14.7$ for the $(1,1)$ soliton and $g=7.05$ for both the $(1,0)$
and $(0,1)$ ones; in the $(\psi,\chi)$ coordinates they are $g=7$
for the $(1,1)$ soliton, $g=16.2$ for the $(0,1)$ one and $g=14.7$
for the $(1,0)$ one.
When the instability sets in, what we observe is that the soliton --
i.e. the discrete configuration smoothly interpolating between the
boundary values -- breaks down and we have instead a configuration
in which all angles are equal to the left boundary value for $n
\le n_0$, and to the right boundary value for $n>n_0$.

In other words, the transition between the two boundary value does
not take place over a (more and less extended) range of sites, but
abruptly at a given site -- which we denoted above as $n_0$, but
of course can be any site. This also shows that in this case we
have a strong degeneration of the Hamiltonian also in the finite
length case (for infinite chains, this is always the case as the
Hamiltonian is invariant under translations), which will show up
in a computational instability for the numerical energy
minimization.

In Fig. \ref{Yak} we show the profiles of the \Y solitons we have
obtained for $g=23.4$ (this is the value corresponding to the
coupling constants in use in our model, see Eq. (\ref{eq:g}) below)
and $g=150$ (this corresponds to the coupling constant value
chosen in \cite{YakPRE}). The profiles we obtain are qualitatively
identical to the inhomogeneous ones presented in
\cite{YakPRE}. Incidentally, we noticed a peculiarly
strong dependence on the initial conditions for the $(1,1)$ mode,
so that those profiles change considerably depending on the
approximation chosen for $2\pi$: e.g., in Fig.~\ref{Yak}c we used
an approximation extremely precise while in Fig.~\ref{Yak}d we
used $2\pi=6.28$. We believe that the first one represents the
correct solution, e.g. also because it is the only one of the two
that respects the symmetry of the equations.
We also produced profiles for the solitons of the \Y Hamiltonian
with respect to the angles $\psi,\chi$ (which
are the coordinates we use in our Hamiltonian). The results are
show in Fig. \ref{Yak2}; in these new coordinates no strong
dependence on the initial conditions was spotted.

\subsection{Solitons in the composite  model}

Let us now turn to the numerical investigation of our  model. We
will consider the case when the intrapair distance at the
equilibrium is zero, i.e we will set $a=r+d_{h}$ (contact
approximation).
Notice that we are not considering the zero-radius approximation
for the bases, so that in general $r\neq d_{h}$.
The Hamiltonian of the system  can be easily derived from the
Lagrangian (\ref{lag}) and is given by \beq H \ = \ T_B + T_b +
V_t + V_s + V_p + V_h + V_w \ . \feq
We use the shorthand notation
$$ \begin{array}{l}
\psip_n = \theta^+_n \ , \ \psim=\theta^-_n \ , \ \Omegap_n =
\phi^+_n \ , \ \Omegam_n=\phi^-_n \ ; \\
\dpsip = \psip_{n+1} - \psi_n \ , \ \spsip = \psip_{n+1}+\psi_n \
; \end{array} $$ and similarly for the other variables; we also
write $$ \alpha = R/r \ , \ \beta = R / d_h \ ; $$ with the
values given in table \ref{tab:cynParms}, it results $\a = 0.92$,
$\b =0.53$. With these notations, we have
\begin{equation}
\label{eq:GaetaHam}
  \begin{array}{rl}
    T_B=&\sum_n I_B\left(\dpsip^2+\dpsim^2\right)\\
    T_b=&\sum_n I_b    [
      \dOmegap^2+\dOmegam^2+\dpsip^2+\dpsim^2+2\dpsip\dOmegap+2\dpsim\dOmegam +\alpha^2(\dpsip^2+\dpsim^2)
      \\
      &+2\alpha(\dpsip^2+\dpsim^2+\dpsip\dOmegap+\dpsim\dOmegam)\cos\Omegap_n\cos\Omegam_n\\
      &+2\alpha(2\dpsip\dpsim+\dpsim\dOmegap+\dpsip\dOmegam)\sin\Omegap_n\sin\Omegam_n
    ]\\
    V_t=&2K_t \sum_n [\cos\Delta_n\psip\cos\Delta_n\psim-1]\\
    V_s=&\frac{1}{2} K_s r^2\sum_n
    4[ 1 + \alpha^2 - \alpha^2\cos(\dpsip)\cos(\dpsim) -
    \cos(\dpsim + \dOmegam)\cos(\dpsip+\dOmegap)\\
      &+2\alpha\cos((\sOmegap-\sOmegam)/2)\sin((\dpsip-\dpsim)/2)
      \sin((\dpsip-\dpsim+\dOmegap-\dOmegam)/2)\\
      &+2\alpha\cos((\sOmegap+\sOmegam)/2)\sin((\dpsip+\dpsim)/2)
      \sin((\dpsip+\dpsim+\dOmegap+\dOmegam)/2)
    ]\\
    V_p=&\frac{1}{2} K_p d_h^2 \sum_n 4[
      (1+\beta)^2+\cos^2(\psim_n+\Omegam_n)+
      2\beta\cos\psim_n\cos\Omegap_n\cos(\psim_n+\Omegam_n)+\beta^2\cos^2\psim_n\\
      &-2\beta(1+\beta)\cos\psip_n\cos\psim_n -2(1+\beta)\cos(\psip_n+\Omegap_n)
      \cos(\psim_n+\Omegam_n)]\\
    V_h=&K_h\sum_n [2 - \cos(\psip_{n+5}-\psim_n) -
    \cos(\psim_{n+5}-\psip_n)]\\
    V_w=&K_w\sum_n [\tanh(\Omegap_n+\Omegam_n)+\tanh(\Omegap_n-\Omegam_n)]
  \end{array}
\end{equation}

Note that we have inserted in the Hamiltonian the confining
potential $V_{w}$ in order to implement dynamically the constraint
(\ref{bc}) for the non-topological angles $\phi^{(a)}$. Adding
this term in the potential is also instrumental in stabilizing the
numerical minimizations.

The Hamiltonian (\ref{eq:GaetaHam}) reduces to that of the \Y
model setting $\Omegap=\Omegam=0$ (and disregarding the helicoidal
term); with this we get
\begin{equation}
  \label{eq:GaetaYakHam}
  \begin{array}{rl}
    T_B=&I_B\sum_n \left(\dpsip^2+\dpsim^2\right)\cr
    T_b=&I_b(1+\alpha^2)\sum_n [\dpsip^2+\dpsim^2]\cr
    V_t=&2K_t\sum_n [\cos\dpsip\cos\dpsim-1]\cr
    V_s=&\frac{1}{2} K_s r^2\sum_n 4(1+\alpha)^2 [1-\cos\dpsip\cos\dpsim]\cr
    V_p=&\frac{1}{2} K_p d_h^2\sum_n  4(1+\beta)^2
    [1-2\cos\psip_n\cos\psim_n +\cos^2\psim_n]\cr
    V_w=&0;
  \end{array}
\end{equation}
Also note that in this case $V_t$ and $V_s$ differ just by a
multiplicative function.

\begin{figure}
  \includegraphics[width=150pt]{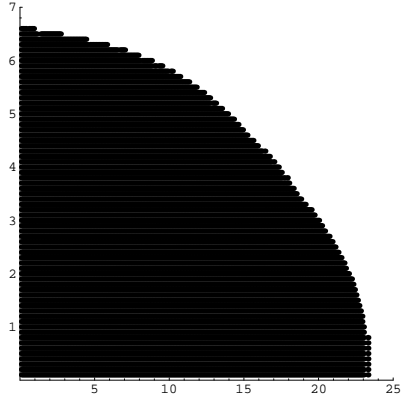}
  \caption{Region of instability (black region) for the discrete solitons
  of topological type
  $(0,1)$.
    in the $(g_t,g_s)$ plane.}
  \label{fig:unstReg}
\end{figure}

The typical energies involved in the different interactions are
given by the coefficients  in Eq. (\ref{eq:GaetaHam}),
$E_{t}=2K_{t},\, E_{s}= \frac{1}{2} K_{s}r^{2},\, E_{p}=
\frac{1}{2} K_{p}d_{h} ^{2},\,E_{h}=K_{h},\, E_{w}=K_{w}$,  which
represent, respectively the typical torsional, stacking, pairing,
helicoidal and confining energies. Using the values of the
physical parameters given in the tables \ref{tab:cynParms},
\ref{tab:dynParms}, and choosing $E_{w}$ in order to keep the
confining energy at least a full order of magnitude smaller than
any other one, we get for the typical interaction energies the
values given in table \ref{tab:energies}.

In order to work with dimensionless quantities, throughout this
section we will measure energies in terms of $E_p=(1/2) K_p
d_h^2=4.0\cdot10^2 {\rm KJ/mol}$. Using the values of the
kinematical parameters given in \ref{tab:para} and those of the
dynamical parameters given by table \ref{tab:dynParms}, the
dimensionless coupling constants turn out to be \beq\lb{nc}
\begin{array}{l}
g_t = E_t/E_p = 0.65 \ , \
g_s = E_s/E_p = 7.2 \ , \ g_p = 1 \ , \\
g_h = E_h/E_p = 0.026 \ , \ g_w = E_w /E_p = 0.001 \ .
\end{array} \feq

\begin{table}
  \centering
  \begin{tabular}{|c|c|c|c|c|}
  \hline
    $E_t$ & $E_s$ & $E_p$ & $E_h$&$E_{w}$ \\
  \hline
    $2.6\cdot10^2 KJ/mol$ & $2.9\cdot10^3 KJ/mol$ & $4\cdot10^2KJ/mol$ & $10
    KJ/mol$ &$4\cdot10^{-1}KJ/mol$ \\
  \hline
\end{tabular}
  \caption{Values of the typical energies characterizing the different
  interactions in the Hamiltonian of Eq.~\ref{eq:GaetaHam}}
  \label{tab:energies}
\end{table}

Note that Eq. (\ref{eq:GaetaYakHam}) implies that, in the limit
$\Omegap=\Omegam=0$, the \Y couplings ($K,g$) and our coupling
constants are related by \beq K \ = \ 2 (1+\beta)^2 \, g_p \ , \ \
g \ = \ 2 \, g_t \ + \ 4 (1+\alpha)^2 \, g_s \ ; \feq this also
yields
\begin{equation}
\label{eq:g}
\frac{g}{K} \ = \ \frac{g_t+2(1+\alpha)^2g_s}{(1+\beta)^2 g_p}
\ \simeq \ \frac{g_t+7.4g_s}{2.3g_p} \ \simeq \ 23 \ .
\end{equation}

Most of the statements made in the previous section for the \Y
Hamiltonian apply almost verbatim  to our case. We obtain an
approximate profile of a soliton, subject to the  boundary
conditions (\ref{fes}), by
minimizing numerically the Hamiltonian through the
``conjugate-gradient'' algorithm, in particular through its
implementations in the GSL \cite{gsl} and in the Numerical Recipes
\cite{NR}.

\begin{figure}
$$\begin{array}{cc}
    \includegraphics[width=150pt]{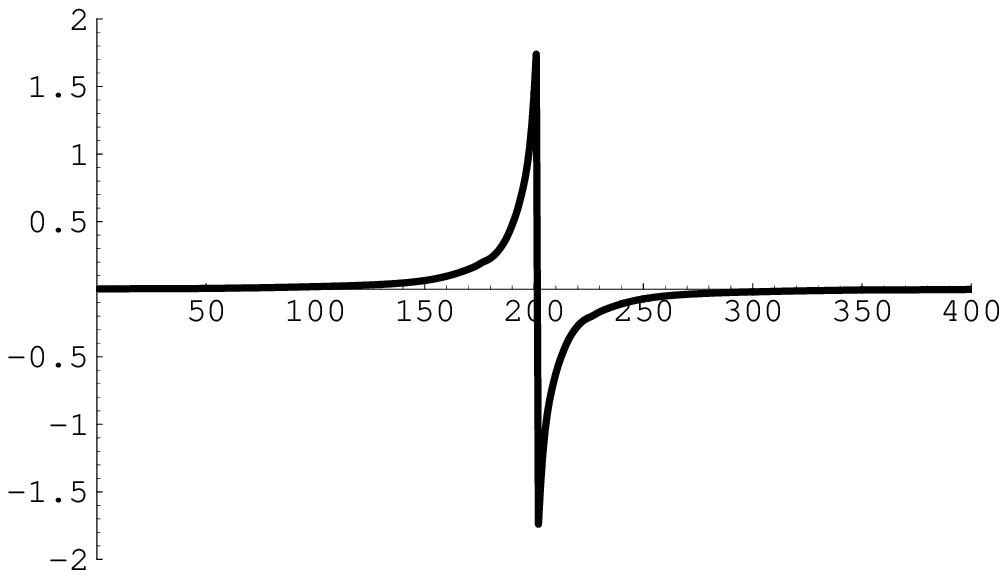}&
    \includegraphics[width=150pt]{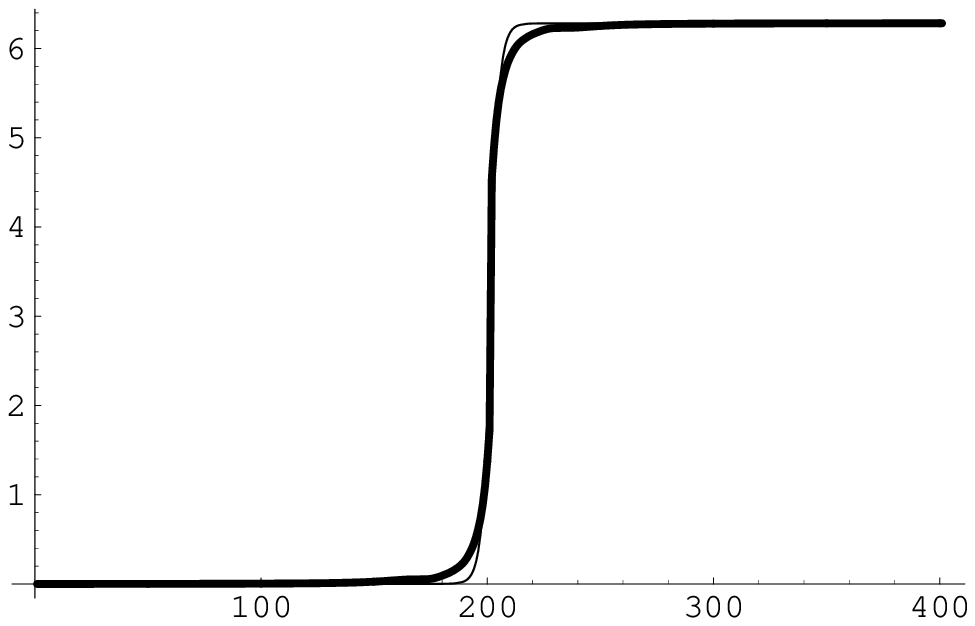}\\
    a&b\\
    \includegraphics[width=150pt]{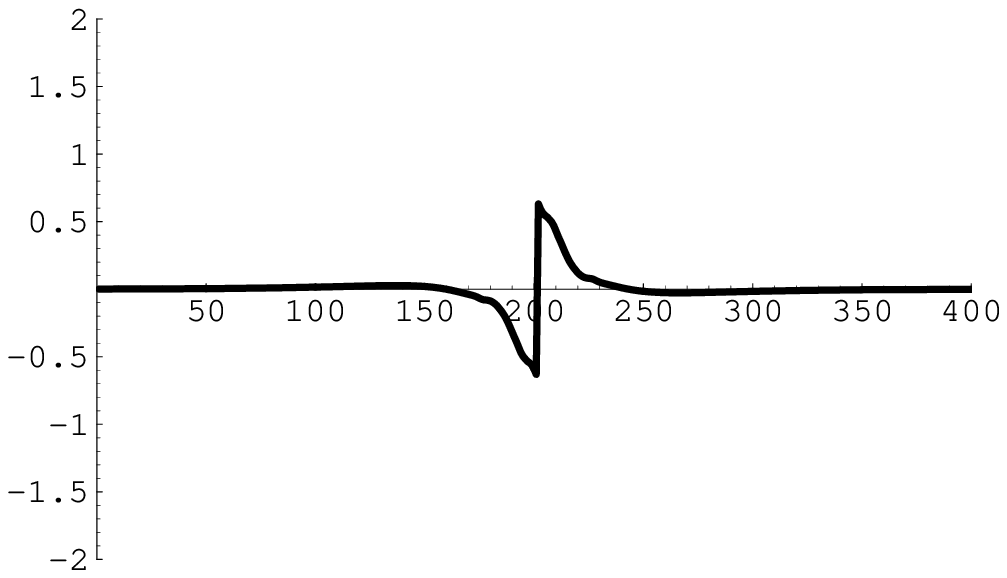}&
    \includegraphics[width=150pt]{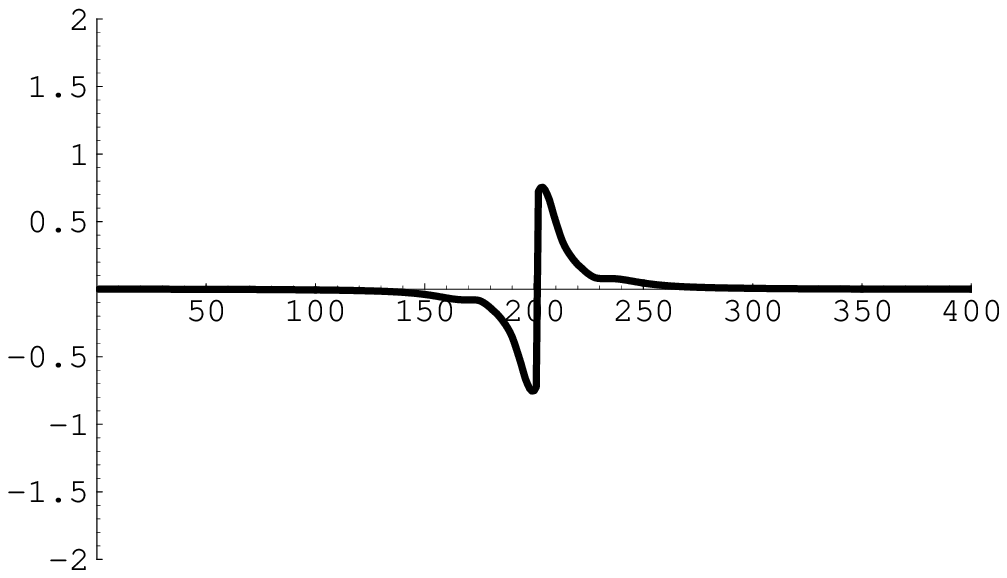}\\
    c&d\end{array} $$
  \caption{Stationary solitonic solutions of our model with energy
  $E=80.06 \,E_{p} $ of topological numbers $(1,0)$ (thick line) compared
  with the solitonic solutions of the \Y model  of energy  $E=75.56 \, E_{p}$
  (thin line).
  Upper left (a): the angle $\psip$; upper right (b): the angle $\psim$;
  lower left (c): the angle $\Omegap$; lower right (d): the angle $\Omegam$.
  The thin line segments visible in (b) show the small difference between
  the profile of the  solitons of our and  of the \Y model.
  }
  \label{fig:g1.0}
\end{figure}

To enforce a particular topological type $(p,q)$ for the soliton
under study we fix the angles at the extremes of the chain so that
$\psip_{-\infty}=\psim_{-\infty}=0$ and $\psip_{+\infty}=2\pi p$,
$\psim_{+\infty}=2\pi q$, while the non-topological angles are
requested to be identically zero at the extremes.
As ``starting point'' for the the algorithm (see above) we use the
natural choice \cite{YakPRE} \beq \begin{array}{l} \psip_n = \pi
 p(1+\tanh(\beta (2n-N))) \ , \\
 \psim_n= \pi q(1+\tanh(\beta (2n-N))) \ , \\
\Omegap_n=\Omegam_n=0 \ , \end{array} \feq where $\beta$ is a
parameter used to adjust the profile of the initial configuration
(the starting point) and $N$ is the number of sites on the chain.
The number $N$ is of course much smaller than in real DNA (usually
$N=4000$ in our simulations) but big enough to ensure that $\psip$
and $\psim$ are constant at the beginning and the end of the chain
within the numerical precision of our computations.

\begin{figure}
  {} \hfill \includegraphics[width=150pt]{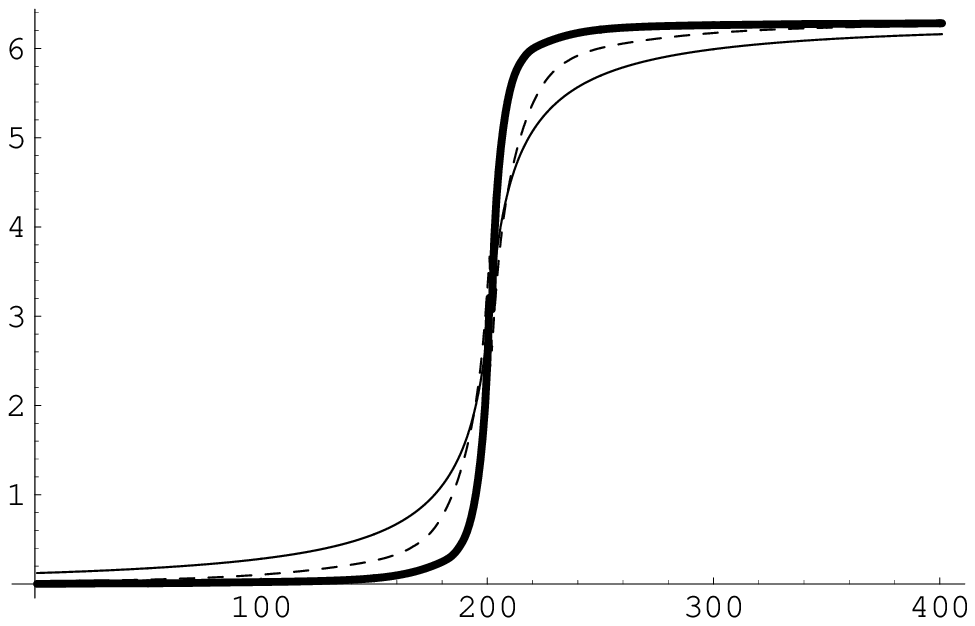}
  \hfill \includegraphics[width=150pt]{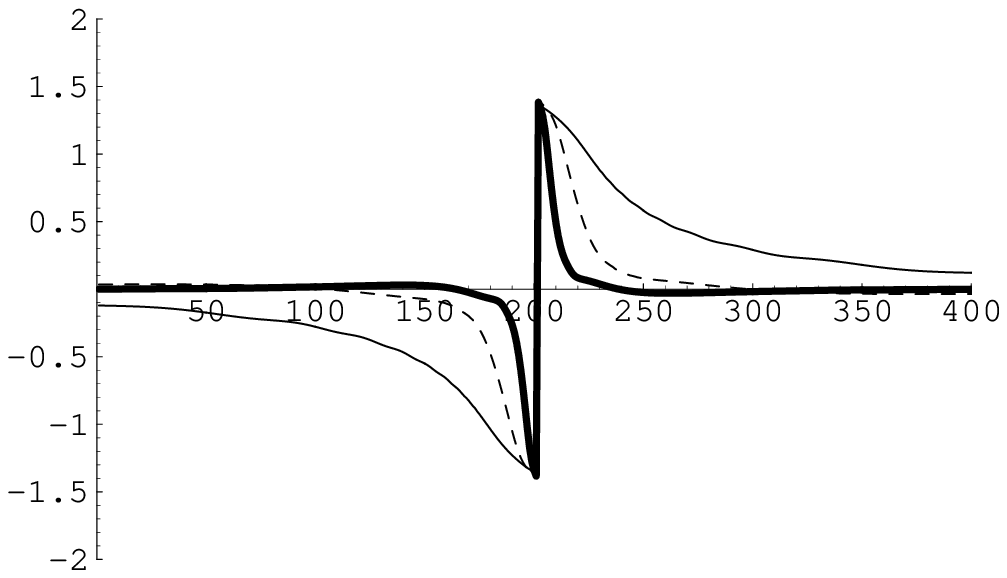} \hfill {}
  \caption{Stationary solitonic solutions of our model with
  topological numbers $(0,1)$ for different values of the normalized
  couplings.  The  angle $\psip$ is depicted
  on the left whereas  $\Omegap$ is depicted
    on the right.
    The soliton relative to the physical coupling constants with energy
    $E=189.9E_{p} $ (thick line)
    is shown together
    with those relative to the coupling constants $g_t=0$, $g_s=46$
   with energy $E=492.4 E_{p} $
    (thin dashed line) and
    $g_t=345$, $g_s=0$ (thin line, $E=388.5 E_{p} $)
    to show how the profile would change at the increasing of the
    coupling constants in the two extreme cases of negligible torsional or stacking interactions.}
  \label{fig:g0.1}
\end{figure}

Like in previous case, there is a threshold for the coupling
constants that must be surpassed for the solitons to be stable; in
Fig. \ref{fig:unstReg} we show the region of instability for
solitons in the $(g_t,g_s)$ plane when we fix the values of the
other coupling constants to be $g_h=0.026$, $g_p=1$, $g_w=0.001$.
As above, in the case of solitons of topological type $(1,0)$ and
$(0,1)$ we always reach the same minimum -- within $10^{-5}$ in
the energy and $10^{-2}$ in the angle -- while $\beta$ varies
across almost two orders of magnitude, provided that $\beta \geq
4$ to avoid falling on the step solution.

The $(1,1)$ soliton also shows the very same behavior as for the
\Y Hamiltonian, namely a different local extrema for every value
of $\beta$ except for a short interval $4 \leq \beta \leq 5.5$
 and for a range of energies wider -- within 2\% --
about 160K; in
the latter we get the same stable behavior observed for the
$(1,0)$ and $(0,1)$ solitons.
In this case again, as in the \Y case, the sensitivity of the
numerical solitonic solutions to the values of the parameter
$\beta$ could be an indication of the existence of many, almost
degenerate, solitonic solution for topological numbers different
than (1,0) or (0,1).
We would like to stress that that the existence of different,
almost degenerate, local minima is a typical feature of most
bio-physical systems. However, our results concerning these point
have to  be regarded just has an indication. Further
investigations, in particular at the analytical level, are needed
in order to draw a definite conclusion.

In Fig. \ref{fig:g1.0} we plot the $(1,0)$ soliton of our model with
the  physical values of the normalized coupling constants
given by Eq. (\ref{nc}) and compare them with those obtained for
the \Y model. The profiles of the topological angles change very
little from the corresponding profiles of the \Y solitons.

%\subsubsection*{Dependence on the parameters}

We have also investigated the deformation of soliton profiles when
adjusting selected parameters of our Hamiltonian.
First, in order to see how the shape of the soliton changes upon
increasing the strength of the torsional/stacking interactions, in
Fig.~\ref{fig:g0.1} we compare the profiles of the $(0,1)$
solitons with profiles corresponding to $g/K\simeq150$ (see
Eq. (\ref{eq:g}), i.e. the coupling constant used in \cite{YakPRE}.

As it is not completely clear how to separate the interaction
strength between torsional and stacking interactions (for our
choice of physical constants in Sect. \ref{s5}) we have  used about
the smallest reasonable value for $K_t$. We present
the profiles corresponding to the two extreme possibilities: the
one in which we put all the strength in the backbone torsional
interaction ($g_t=345$, $g_s=0$), and the one in which we put all
of it in the bases stacking ($g_t=0$, $g_s=46$). The effect is the
 widening of both the soliton and the non-topological
profiles by roughly a factor 4 in the first case and of a factor 2
in the second case.

\begin{figure}
$$\begin{array}{cc}
 \includegraphics[width=150pt]{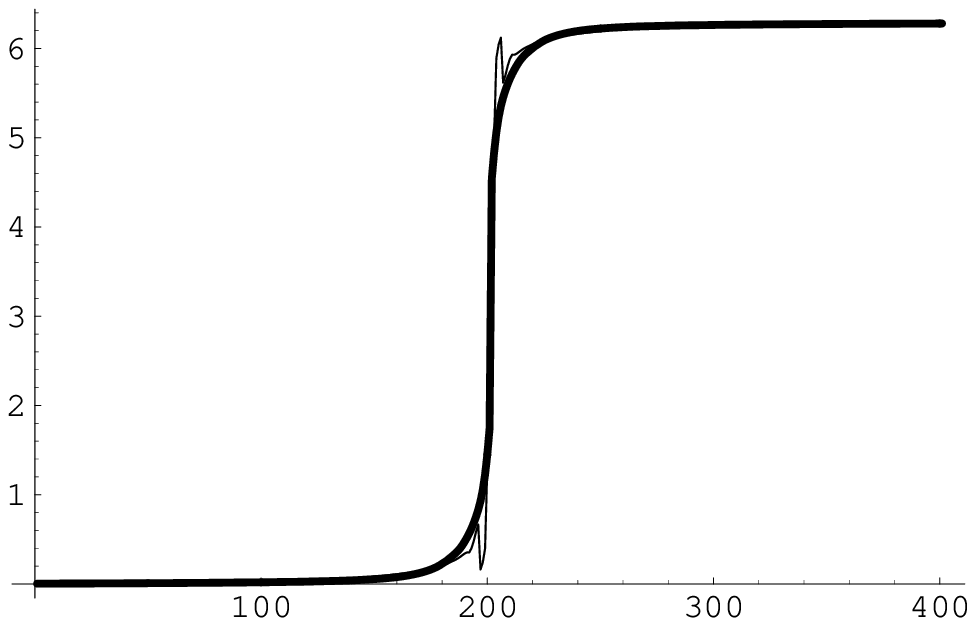}&
 \includegraphics[width=150pt]{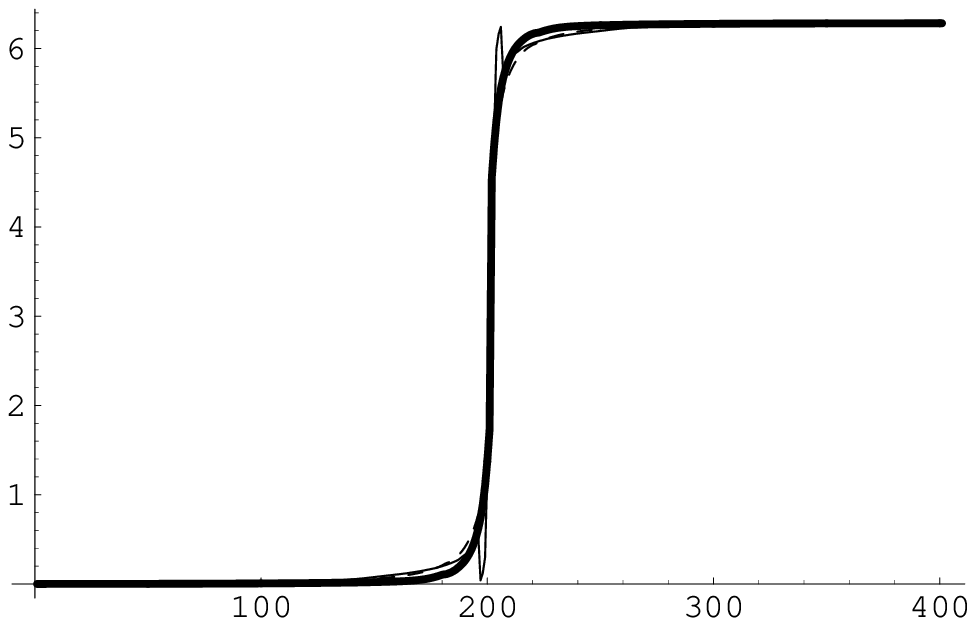}\\
    a&b\\
    \includegraphics[width=150pt]{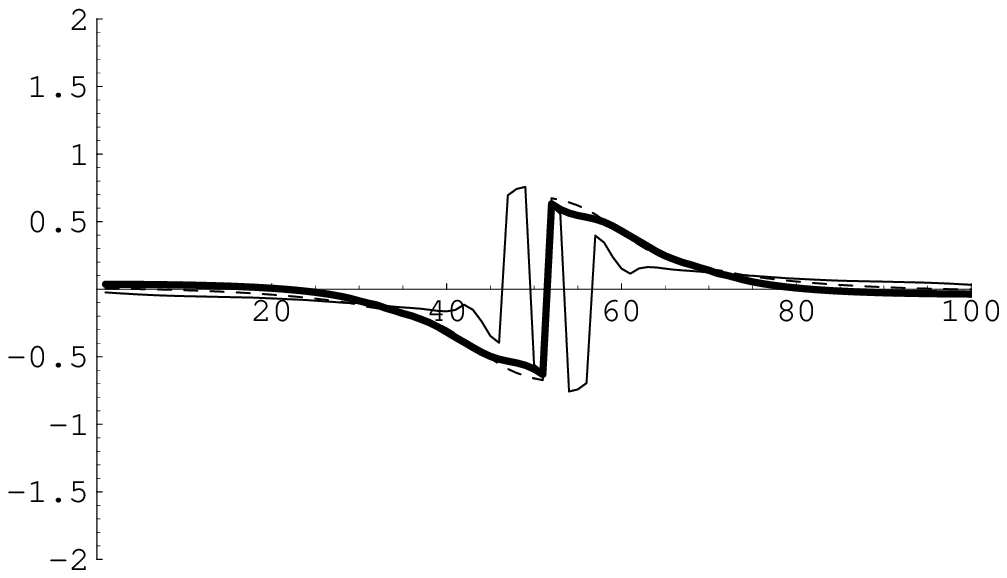}&
    \includegraphics[width=150pt]{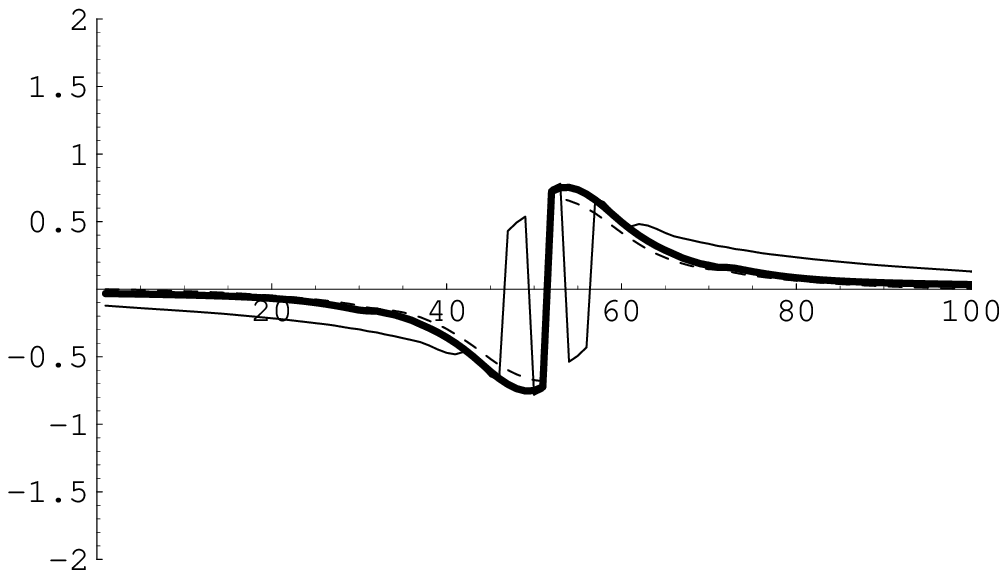}\\
    c&d\end{array} $$
  \caption{Comparison of stationary solitonic solutions  of our
  model  with those obtained using a modified pairing potential
  $V_{p}$. Upper left (a): the angle $\psip$; upper right (b): the angle $\psim$;
  lower left (c): the angle $\Omegap$; lower right (d): the angle $\Omegam$.
  The thick line represent a soliton with $E=80.06$ and topological numbers $(1,1)$).
  The thin dashed line gives the profile for the same soliton with
  energy $E=74.41 E_{p} $ and
  with the ``correct''
  pairing potential $V_p=g_p(\rho-d_0/d_h)^2$. The latter solitonic
  solution has been derived taking $d_0=3.2d_h$,
  namely a order of magnitude bigger than its physical value, to enhance the
  profile differences (an almost identical profile is obtained if we suppress
  the helicoidal term from the Hamiltonian). The thin continuos line
  ($E=102.9 E_{p} $) is the
  profile we get by increasing the helicoidal term to $g_h=1$.}
  \label{fig:g1.1}
\end{figure}

In Fig.~\ref{fig:g1.1} we compare the profiles of the soliton
$(1,1)$ with those obtained by using the correct distance function
for $V_p$, namely by replacing $g_p\rho^2$ with
$g_p(\rho-d_0/d_h)^2$ (where $d_0 \simeq 2~\AA$ is the equilibrium
distance between two bases in a pair \footnote{In order to avoid
any confusion, we stress that here the ``distance'' between bases
refers e.g. to the N--H distance in a N--H--O hydrogen bond; the
total length of the bond is about $3 \AA$, and often one refers to
this as the interbase distance. Here instead we consider the $H$
atom -- which lies at about $1 \AA$ from the nearer atom in the H
bond -- as part of one of the bases and hence consider the
distance of it from the other atom as the interbase distance.}),
and by varying the helicoidal interaction term. No relevant
changes are detected in the first case: relative differences in
energies and angles are of the order of $10^{-2}$ in energy and
$10^{-1}$ in the angles; even increasing the base-pairs distances
by two orders of magnitude these results do not  modify the
situation.

As for the helicoidal term, we get variations of the same order of
magnitude as above if we simply turn it off. If we instead
increase the coupling constant by one order of magnitude
($g_h=0.26$), then we get energy and angles changes of the order
of $10^{-1}$ and by increasing it to $g_h=1$ we arrive to changes
of the order of $10^0$ in both the angles and the energy.

Raising $g_h$ up to $g_h=2$ leads to the disappearance of the
soliton; it seems reasonable to argue that this is due to such an
interaction favoring a sharper transition between limit behaviors,
so that the discreteness effect discussed in the previous
subsection arises.

\subsection{Discussion}

The numerical analysis we have performed shows  the
existence of solitonic solutions of our composite DNA model.
The profiles of the topological solitons -- in particular, the
part relating to the topological degree of freedom -- of our model
are both qualitatively and quantitatively very similar to those of
the Y model. This means that the most relevant (for DNA
transcription) and characterizing feature of the nonlinear DNA
dynamics present in the Y model is preserved by considering
geometrically more complex and hence more realistic DNA models.

Moreover, the topological soliton profiles of our model seem to
change very little when either the physical parameters change in a
reasonable range or also the form of the potential modelling the
pairing interaction is modified to a more realistic form.
In particular, the form of the topological solitons are very
little sensitive to the interchange of torsional and stacking
coupling constant.

This feature add other reasons why the Y model, although based on
a strong simplification of the DNA geometry, works quite well in
describing solitonic excitations. The Y model, indeed, does not
distinguish between torsional and stacking interaction; but, as we
have shown, this distinction is not relevant -- at least as long
as one is only interested in the existence and form of the soliton
solutions.
The ``compositeness'' of our model becomes relevant -- and
rather crucial -- when it comes on the one hand to allowing the
existence of solitons {\it together} with requiring a physically
realistic choice of the physical parameters characterizing the
DNA, and on the other hand to have also predictions fitting
experimental observations for what concerns quantities related to
small amplitude dynamics, such as transverse phonons speed. In
other words, the somewhat more detailed description of DNA
dynamics provided by our model allows it to be effective -- with
the same parameters -- across regimes, and provide meaningful
quantities in both the linear and the fully nonlinear regime.

The solitonic solutions of the composite model share also two
other features with those of the Y model, namely the presence of a
numerical instability and the existence of quasi-degenerate
solutions for solitonic configurations with higher topological
numbers.

We expect that the model considered here is the simplest DNA model
describing rotational degrees of freedom which, with physically
realistic values of the coupling constants and other parameters,
allows for the existence of topological solitons and at the same
time is also compatible with observed values of bound energies and
phonon speeds in DNA.

\section{Summary and conclusions}
\lb{s10}

Let us, in the end, summarize our discussion and the results of
our work, and state the conclusions which can be drawn from it.

\subsection{Summary and results}

Following the work by Englander {\it et al.} \cite{Eng}, different
authors have considered simple models of the DNA double chain --
focusing on rotational degrees of freedom -- able to support
dynamical and topological solitons \cite{YakuBook}, supposedly
related to the transcription bubbles present in real DNA and
playing a key role in the transcription process.

These models usually consider a single (rotational) degree of
freedom per nucleotide \cite{YakuBook}, albeit models with one
rotational and one radial degree of freedom per nucleotide have
also been considered \cite{BCP,BCPR,CM,Joy,YakSBP} (as an
extension of ``purely radial'' models \cite{PB,PeyNLN}, considered
in the study of DNA denaturation). A simple model which has been
studied in depth is the so called Y model \cite{YakPLA}. This
supports topological solitons (of sine-Gordon type) and provides
correct orders of magnitude for several physically relevant
quantities \cite{GRPD}; on the other hand, the soliton speed
remains essentially a free parameter \cite{GaeSpeed,GaeJBP}, and
the speed of transverse phonons can be made to have a physical
value only by assigning unphysical values to the coupling
constants of the model \cite{YakPRE}.

Here we have considered an extension of the Y model, with two
degrees of freedom -- both rotational -- per nucleotide; one of
these is associated to rotations of the nucleoside (unit of the
backbone) around the phosphodiester chain and is topological --
i.e. can go round the $S^1$ circle -- while the other is
associated to rotations of the attached nitrogen base around the
$C_1$ atom in the sugar ring, and due to sterical hindrances is
non-topological -- i.e. rotations are limited to a relatively
small range around the equilibrium position. We denoted this as a
``composite Y model''.

Several parameters appear in the model; some of these are related
to the geometry and the kinematics of the DNA molecule, while
other are coupling constants entering in the potential used to
model intramolecular interactions. We have assigned values to the
first kind of parameters following from available direct
experimental observations, and for the second kind of parameters
we used experimental data on the ionization energies of the
concerned couplings and the form of the potentials appearing in
the model. That is, these parameters were {\it not}  chosen by
fitting dynamical predictions of the model; see sect. V for
detail.
We have first considered small amplitude dynamics (sect. VI); this
yields the dispersion relations and produced some prediction on
the phonon speed and the optical frequency for the different branches.
These prediction are
a first success of the present model, in that it was shown in
sect. VI that taking the physical order of magnitude for the pairing,
stacking, torsional and helicoidal interaction, one can obtain the order of
magnitude of the
experimentally observed speed for transverse phonon excitations and
the frequency threshold for the optical branch. In particular, using
a physical value for the stacking and torsional interaction energy we
get  a value for the transverse phonon speed which is about three times the
``correct'' one.
It should be considered, in looking at this
value, that we modelled the intrapair interaction by a very simple
and non realistic potential (with the aim of both keeping
computations simple and allowing direct comparison with the
standard Y model by making the same simplifying assumptions as
there). As for comparison with standard Y model, hence for an
evaluation of the advantages brought by considering a more
articulated geometry of the nucleotide, it should be recalled that
the numerical computations of Yakushevich, Savin and Manevitch
\cite{YakPRE} (which we repeated, and fully confirmed) show in
order to obtain the experimentally observed speed for transverse
phonon excitations in the framework of the standard Y model, one
should take a coupling constant for the transverse intrapair
interaction which is about 6000 times the ``correct'' one.

We passed then to consider the fully nonlinear regime, and in
particular to look for solitonic-like travelling excitations.
These should have smooth variations on the space scale of
nucleotides, hence we passed to a continuum description and field
equations; by using the chain exchange symmetry, we considered
fully symmetric and antisymmetric reductions, see (\ref{6.6}) and
(\ref{6.8}). By a travelling wave ansatz we reduced these to a
system of two coupled second order ODEs for $\vth (z)$ and $\vphi
(z)$, see (\ref{6.12}) and (\ref{6.13}). Here $\vth$ is the
topological angle, i.e. the variable associated to the topological
field, and $\vphi$ is the non topological angle, i.e. the variable
associated to the non topological field.

The finite energy condition (\ref{finkin}), (\ref{finpot})
requires that the solutions to this system of ODEs satisfy certain
limit conditions (see (\ref{fes} ). These in turn imply that solutions
satisfying them can be classified according to two topological
indices (winding numbers for the topological fields; in the
symmetric or antisymmetric case, one index is enough to determine
the other as well).

We have also shown that the standard Y model can be obtained from
the composite Y model by a limiting procedure (sect. VII); this
also reduces the solitons of the composite model to solitons of
the Y model. However, the limiting procedure requires that a
certain condition is satisfied, see eq. (\ref{8.5}), and this in
turn constraints the speed of solitonic excitations; see
(\ref{8.6}). Thus, the requirement to obtain the standard Y
solitons in a certain limit fixes the speed of solitons; the
resulting speed is just the speed of long waves as determined by
the dispersion relations.

Finally, in sect. IX we conducted a careful numerical
investigation of the simpler soliton solutions for the composite Y
model. We preliminarily checked our numerical routines on the
standard Y model and fully confirmed the results of Yakushevich,
Savin and Manevitch \cite{YakPRE}, also confirming certain
instability phenomena they reported. We then considered the
solitons for the composite Y model with the value of parameters
descending from their physical meaning (i.e. with no parameter
fitting), confirming their existence, properties and stability. We
also showed how the profile of the soliton component corresponding
to the topological degree of freedom is extremely similar to the
standard Y soliton with same topological numbers. We considered
next the stability of these soliton solutions upon varying the
parameters of the model, and observed that as in the standard Y
case there is a stability threshold. Thus, the existence and
stability of soliton solutions for physical values of the
parameters is a nontrivial prediction.

\subsection{Discussion and conclusions}

The composite Y model considered here retains all the favorable
features of the standard Y model. At the same time, its more
articulated geometry allows at the same time -- and with physical
values of the coupling constants and other parameters entering in
the model -- to reproduce relevant value of physical quantities
related to the linear regime (such as speed of transverse phonon,
which was a critical test for the standard Y model) and support
stable soliton solutions.

Further, and at difference with the standard Y model, it provides
a precise prediction for the soliton speed; this is quite
reasonable physically, as it corresponds to the speed of long
waves as obtained from the dispersion relations for the model.
Thus, our model passed some -- in our opinion, significant --
quantitative test and provides precise predictions.

It should also be stressed that we used -- both to simplify the
mathematics and to have a direct comparison with the standard Y
model -- a very simple form for the intrapair coupling potential
(and at some stage also resorted to the ''contact approximation''
to get simpler formulas, again as in the standard Y model
treatment). It is quite conceivable that adopting a more realistic
potential will provide better estimates of relevant physical
quantities, in particular for quantities related to the linear
regime. However, experience recently gained with the standard Y
model \cite{GaeY1,GaeY2} suggests that the predictions related to
the fully nonlinear regime are rather little sensitive to the
detailed form of the potential and to adopting or otherwise the
contact approximation; we are thus rather confident that future
work with more realistic potentials will confirm the results
obtained in the simple setting considered here.

Finally, we would like to remark a very relevant feature of our
model. All the DNA models amenable to analytic treatment look at
homogeneous DNA, albeit the genetic information lies precisely in
the non homogeneous part of the DNA (i.e. the base sequence; bases
have rather different physical and geometrical characteristics).
Our discussion was no exception, and we considered identical bases
with ``average'' geometrical and physical characteristics. But,
the degrees of freedom we considered for each nucleotide are one
concerned with the uniform part of the DNA molecule (the
nucleosides), the other with the non homogeneous part (the base
sequence). Moreover, it turned out that -- for what concerns
soliton excitations -- the most relevant role is played by the
(topological) variables associated to the uniform part, which are
directly at play in the topological solitons, while the (non
topological) variables associated to the non uniform part are in a
way just accompanying the soliton.

This suggests that, within the framework of composite models, the
non homogeneous case can be studied as a (non-singular)
perturbation of the homogeneous case; needless to say, by this we
mean an analytical -- albeit approximated -- study, and not just a
numerical one. This represents a significant advance with respect
to what is possible with simple models considered so far.

\section*{Acknowledgements}

This work received support by the Italian MIUR (Ministero
dell'Istruzione, Universit\`a e Ricerca) under the program
COFIN2004, as part of the PRIN project {\it ``Mathematical Models
for DNA Dynamics ($M^2 \times D^2$)''}.

%\vfill\eject

\end{document}